\documentclass[aps,twocolumn,preprintnumbers,amsmath,amssymb,superscriptaddress]{revtex4-1}

\usepackage{graphicx}
\usepackage{dcolumn}
\usepackage{epsfig}
\usepackage{color}
\usepackage{ulem}
\usepackage{bbm}

\usepackage{hyperref,cleveref}

\usepackage{lipsum}
\newcommand{\pmp}{$Pm\bar{3}m1^\prime$}

\newcommand{\ket}[1]{\left|#1\right\rangle}		
\newcommand{\bra}[1]{\left\langle#1\right|}		

\newcommand{\bI}{\mathbbm{1}}

\newcommand{\bee}{\begin{equation}}
\newcommand{\ee}{\end{equation}}
\newcommand{\bma}{\begin{pmatrix}}
\newcommand{\ema}{\end{pmatrix}}
\newcommand{\balig}{\begin{align}}
\newcommand{\ealig}{\end{align}}

\begin{document}

\title{Creating and controlling Dirac fermions, Weyl fermions, and nodal lines \\ in the magnetic antiperovskite Eu\textsubscript{3}PbO}

\author{Moritz M. Hirschmann}
\thanks{These authors contributed equally to this work.}
\affiliation{Max-Planck-Institut f\"{u}r Festk\"{o}rperforschung, Heisenbergstra\ss{}e 1, D-70569 Stuttgart, Germany}

\author{Alexandra S. Gibbs}
\thanks{These authors contributed equally to this work.}
\affiliation{ISIS Pulsed Neutron Facility, STFC, Rutherford Appleton Laboratory, Chilton, Didcot, Oxfordshire, OX11 0QX, United Kingdom}
\affiliation{School of Chemistry, University of St Andrews, North Haugh, St Andrews, Fife, KY16 9ST, United Kingdom}
\thanks{These authors contributed equally to this work.}

\author{Fabio Orlandi}
\affiliation{ISIS Pulsed Neutron Facility, STFC, Rutherford Appleton Laboratory, Chilton, Didcot, Oxfordshire, OX11 0QX, United Kingdom}

\author{Dmitry Khalyavin}
\affiliation{ISIS Pulsed Neutron Facility, STFC, Rutherford Appleton Laboratory, Chilton, Didcot, Oxfordshire, OX11 0QX, United Kingdom}

\author{Pascal Manuel}
\affiliation{ISIS Pulsed Neutron Facility, STFC, Rutherford Appleton Laboratory, Chilton, Didcot, Oxfordshire, OX11 0QX, United Kingdom}

\author{Vahideh Abdolazimi}
\affiliation{Max-Planck-Institut f\"{u}r Festk\"{o}rperforschung, Heisenbergstra\ss{}e 1, D-70569 Stuttgart, Germany}

\author{Alexander Yaresko}
\affiliation{Max-Planck-Institut f\"{u}r Festk\"{o}rperforschung,
Heisenbergstra\ss{}e 1, D-70569 Stuttgart, Germany}

\author{J\"{u}rgen Nuss}
\affiliation{Max-Planck-Institut f\"{u}r Festk\"{o}rperforschung,
Heisenbergstra\ss{}e 1, D-70569 Stuttgart, Germany}

\author{H.~Takagi}
\affiliation{Max-Planck-Institut f\"{u}r Festk\"{o}rperforschung,
Heisenbergstra\ss{}e 1, D-70569 Stuttgart, Germany}
\affiliation{Institute for Functional Matter and Quantum Technologies, University of Stuttgart, Pfaffenwaldring 57, D-70550 Stuttgart, Germany}
\affiliation{Department of Physics, The University of Tokyo, 7-3-1 Hongo, Bunkyo-ku, Tokyo 113-0033, Japan}

\author{Andreas P. Schnyder}
\email{Corresponding author. E-mail: a.schnyder@fkf.mpg.de}
\affiliation{Max-Planck-Institut f\"{u}r Festk\"{o}rperforschung,
Heisenbergstra\ss{}e 1, D-70569 Stuttgart, Germany}

\author{Andreas W. Rost}
\affiliation{Max-Planck-Institut f\"{u}r Festk\"{o}rperforschung,
Heisenbergstra\ss{}e 1, D-70569 Stuttgart, Germany}
\affiliation{Institute for Functional Matter and Quantum Technologies, University of Stuttgart, Pfaffenwaldring 57, D-70550 Stuttgart, Germany}
\affiliation{School of Physics and Astronomy, University of St Andrews, North Haugh, St Andrews, Fife, KY16 9SS, United Kingdom}

\begin{abstract}
The band topology of magnetic semimetals is of interest both from the fundamental science point of view and with respect to potential spintronics and memory applications.  Unfortunately, only a handful of suitable topological semimetals with magnetic order have been discovered so far. One such family that hosts these characteristics is the antiperovskites, $A_{3}B$O, a family of 3D Dirac semimetals. The $A$=Eu$^{2+}$ compounds magnetically order with multiple phases as a function of applied magnetic field. Here, by combining band structure calculations with neutron diffraction and magnetic measurements, we establish the antiperovskite Eu$_3$PbO as a new topological magnetic semimetal. This topological material exhibits a multitude of different topological phases with ordered Eu moments which can be easily controlled by an external magnetic field. The topological phase diagram of Eu$_3$PbO includes an antiferromagnetic Dirac phase, as well as ferro- and ferrimagnetic phases with both Weyl points and nodal lines. For each of these phases, we determine the bulk band dispersions, the surface states, and the topological invariants by means of \textit{ab-initio} and tight-binding calculations.  Our discovery of these topological phases introduces Eu$_3$PbO as a new platform to study and manipulate the interplay of band topology, magnetism, and transport. 
\end{abstract}

\date{\today}
\maketitle

\noindent

With the great success of topological band theory for insulators~\cite{hasan:rmp,qi:rmp,chiu_review_RMP}, recent research efforts
have branched out to study the topological properties of metals and semimetals~\cite{chiu_review_RMP,volovikLectNotes13,armitage_mele_vishwanath_review,burkov_review_2018,yang_ali_review_ndoal_line}. 
In contrast to ordinary semimetals, topological semimetals exhibit Fermi surfaces which are in close proximity to a band degeneracy formed by a crossing
of valence and conduction bands. These band degeneracies are protected by a nontrivial topology of the electronic wave functions and
give  rise to a number of intriguing physical phenomena, such as ultra-high mobility~\cite{liangOngTransportCd3As2,shekhar_yan_NbP_nat_phys_2015}, Fermi arc or drumhead surface states~\cite{xu_hasan_fermi_arc_science_15,bian_hasan_drumhead_PRB_16}, 
unconventional magnetoresistance~\cite{magnetoresistance_ZrSiS_schoop_parkin,wang_zhang_ZrSiS_magnetoresistance,singha_mandal_magnetotransport_ZrSiS}, 
and anomalous transport properties potentially related to quantum anomalies~\cite{huang_dai_chen_chiral_anomaly_TaAs,zhang_hasan_anomaly_nat_commun_16,rui_parity_anomaly_DNLSM_arXiv}.
Examples of topological semimetals include graphene~\cite{castroNetoRMP09}, Dirac and Weyl semimetals~\cite{WanVishwanathSavrasovPRB11,BurkovBalentsPRB11,young_kane_rappe_Dirac_3D_PRL_12}, and nodal-line semimetals~\cite{xie_schoop_Ca3P2_apl_15,nodal_line_Yang,nodal_line_Yamakage,heikkila_volovik_JETP_11}. 
Besides these, there exist magnetic topological semimetals, which combine long-range magnetic order  with
 nontrival band topology \cite{Vergniory2018,bernevig2022progress,bernevig2022progress,Nie2020EuB6,Nie2022K2Mn3As3O12,li2019intrinsic,shekhar2018_RPtBiWeyl,Su_2020_EuCd2Sb2_WeylSM,lee2021evidence,%
 Schoop2018,%
higo_magnetic_weyl_anom_hall_nat_15,
nayak_parkin_science_advances_Mn3Ge,
yang_parkin_yan_NJP_17,
kuroda_nakatsuji_Mn3Sn_Nat_Matt_17,
wang_bernevig_PRL_16,
Puphal2020_CeAlGe,
gaudet2021weylNdAlSi,
Liu2021_Eu2Ir2O7,
schroter2020weyl,
liu2019magnetic,borisenko2019time,%
tang_zhang_nat_phys_2016,
Park_2011_SrMnBi2,
liu2017magnetic}. 
 Unlike their nonmagnetic counterparts, these magnetic topological materials 
are  highly tunable by   external magnetic fields. In magnetic Weyl semimetals, for example, 
one can envision moving and manipulating Weyl points using applied fields, which in 
turn alters their topological transport characteristics. This property
could be exploited for technological applications, e.g., for the 
development of next-generation spintronic devices~\cite{jungwirth_PRL_17}.

Similar to the research on nonmagnetic topological semimetals there have been great efforts in the study of systems where magnetism coexists with band topology
\cite{Vergniory2018,bernevig2022progress,bernevig2022progress,Nie2020EuB6,Nie2022K2Mn3As3O12,li2019intrinsic,shekhar2018_RPtBiWeyl,Su_2020_EuCd2Sb2_WeylSM,lee2021evidence,%
 Schoop2018,%
higo_magnetic_weyl_anom_hall_nat_15,
nayak_parkin_science_advances_Mn3Ge,
yang_parkin_yan_NJP_17,
kuroda_nakatsuji_Mn3Sn_Nat_Matt_17,
wang_bernevig_PRL_16,
Puphal2020_CeAlGe,
gaudet2021weylNdAlSi,
Liu2021_Eu2Ir2O7,
schroter2020weyl,
liu2019magnetic,borisenko2019time,%
tang_zhang_nat_phys_2016,
Park_2011_SrMnBi2,
liu2017magnetic}. 
However, despite their promising potential for applications and many theoretical proposals~\cite{bernevig2022progress,Nie2020EuB6,Nie2022K2Mn3As3O12,li2019intrinsic}, only a handful of promising topological magnetic Weyl semimetals has been 
identified so far~\cite{shekhar2018_RPtBiWeyl,Su_2020_EuCd2Sb2_WeylSM,lee2021evidence}. 
Unfortunately, several magnetic candidate materials for topological semimetals have additional bands at the Fermi energy, which dilutes the topological transport properties~\cite{
Schoop2018,%
higo_magnetic_weyl_anom_hall_nat_15,
nayak_parkin_science_advances_Mn3Ge,
yang_parkin_yan_NJP_17,
kuroda_nakatsuji_Mn3Sn_Nat_Matt_17,
wang_bernevig_PRL_16,
Puphal2020_CeAlGe,
gaudet2021weylNdAlSi,
Liu2021_Eu2Ir2O7,
schroter2020weyl,
liu2019magnetic,borisenko2019time}.
In other cases, antiferromagnetism (AFM) occurs in centrosymmetric materials, which leads to (gapped) Dirac points instead of Weyl points~\cite{
tang_zhang_nat_phys_2016,
Park_2011_SrMnBi2,
liu2017magnetic
}.

\begin{figure}[t!]
\begin{center}
\includegraphics[width=0.5\textwidth]{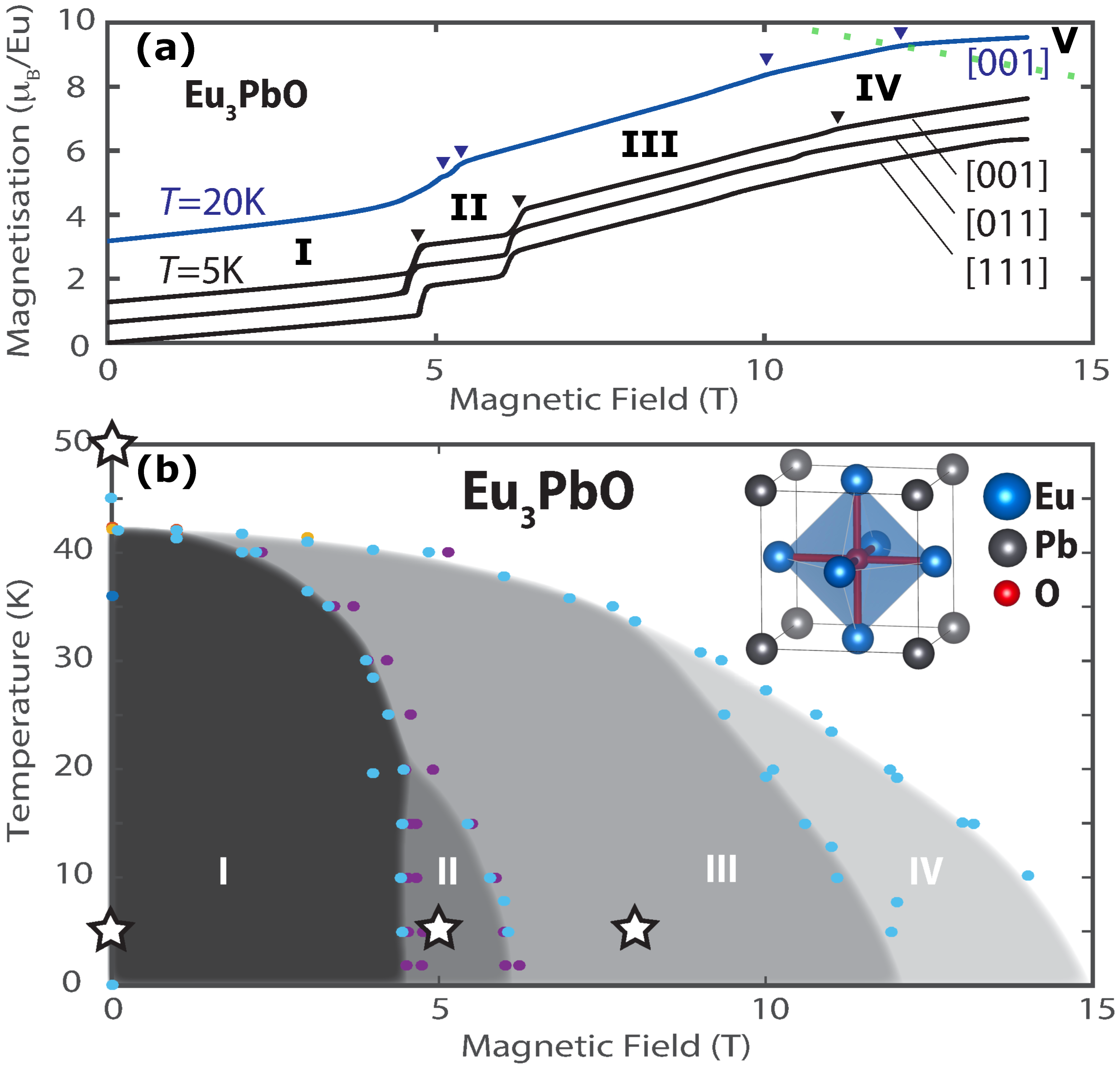}
\end{center}
\caption{\textbf{Magnetic Phase diagram of Eu$_3$PbO.} \label{mFig1}
({\bf a}) Magnetization of a single crystal of Eu$_3$PbO for several applied magnetic field directions at both 5~K and 20~K. The arrows show the location of the phase transitions with the roman numerals denoting the phases. The dashed green line is a guide to the eye indicating the approximate phase boundary between IV and V, the highest field transition to phase V is outside the field range at 5~K, but within it at 20~K. 
({\bf b})~The magnetic phase diagram of Eu$_3$PbO based on magnetization data, both as a function of temperature (purple) and field (blue), and specific heat measurements (orange) on powder samples. Phase I is antiferromagnetic, phases II and III ferrimagnetic, phase IV likely ferrimagnetic and phase V fully polarised. The star-shaped markers indicate the position of our neutron scattering experimental datasets. 
The inset depicts the crystal structure of Eu$_3$PbO.
}
\end{figure}

One fundamental reason for the slow progress in finding ideal magnetic Weyl semimetals
is that these are by their very nature correlated,
which makes their theoretical characterization more demanding.  In fact,  
\textit{ab-initio} calculations in general cannot predict the magnetic structure, which determines not only  the magnetic  symmetries but
also the topological properties of the material. 
Hence, to characterize the topology of magnetic semimetals it is essential to
combine \textit{ab-initio} band structure calculations and  band topology analysis with detailed neutron diffraction measurements determining the magnetic structure. 
Here, we perform such a combined investigation to study the interplay of band topology  with magnetism in the cubic antiperovskite Eu$_3$PbO~\cite{Nuss:dk5032}. 
As a function of magnetic field we observe four magnetically ordered phases:
a noncollinear antiferromagnetic state at zero field, three ferrimagnetic phases at intermediate fields, and a fully polarized ferromagnetic order at large fields (Fig.~\ref{mFig1}).
Using \textit{ab-initio} derived tight-binding models we determine the topological band degeneracies for each of these phases. 
Interestingly, we find that the antiferromagnetic state exhibits  gapped Dirac cones,   while the ferrimagnetic and ferromagnetic phases have Weyl points together with nodal lines.
We derive the topological invariants that protect these band crossings and compute the associated surface states.
Because the Weyl points in the ferri- and ferromagnetic phases act as sinks and sources of Berry flux, these phases
exhibit large Berry curvatures, which enhances the  anomalous  Hall current.

It is remarkable that Eu$_3$PbO realizes such a rich variety of different topological phases, which can be controlled by an applied field.
Therefore, the electronic structure and the band topology of Eu$_3$PbO can be easily tuned and manipulated with an external field.
We demonstrate that this can be achieved in two different ways: 
(i) by rotating the magnetization direction within a given phase 
and (ii) by driving Eu$_3$PbO across magnetic phase transitions.
Through these mechanisms, it is thus possible to tune the Berry curvature of
the valence and conduction bands, which in turn controls the strength of the anomalous Hall current.
Hence, Eu$_3$PbO represents an ideal platform to study the interdependence among magnetism, band topology, and
transport.

 \vspace{0.3cm}
 
\section{Results}

\begin{figure*}[btp]
\centering
\includegraphics[width=0.95\textwidth]{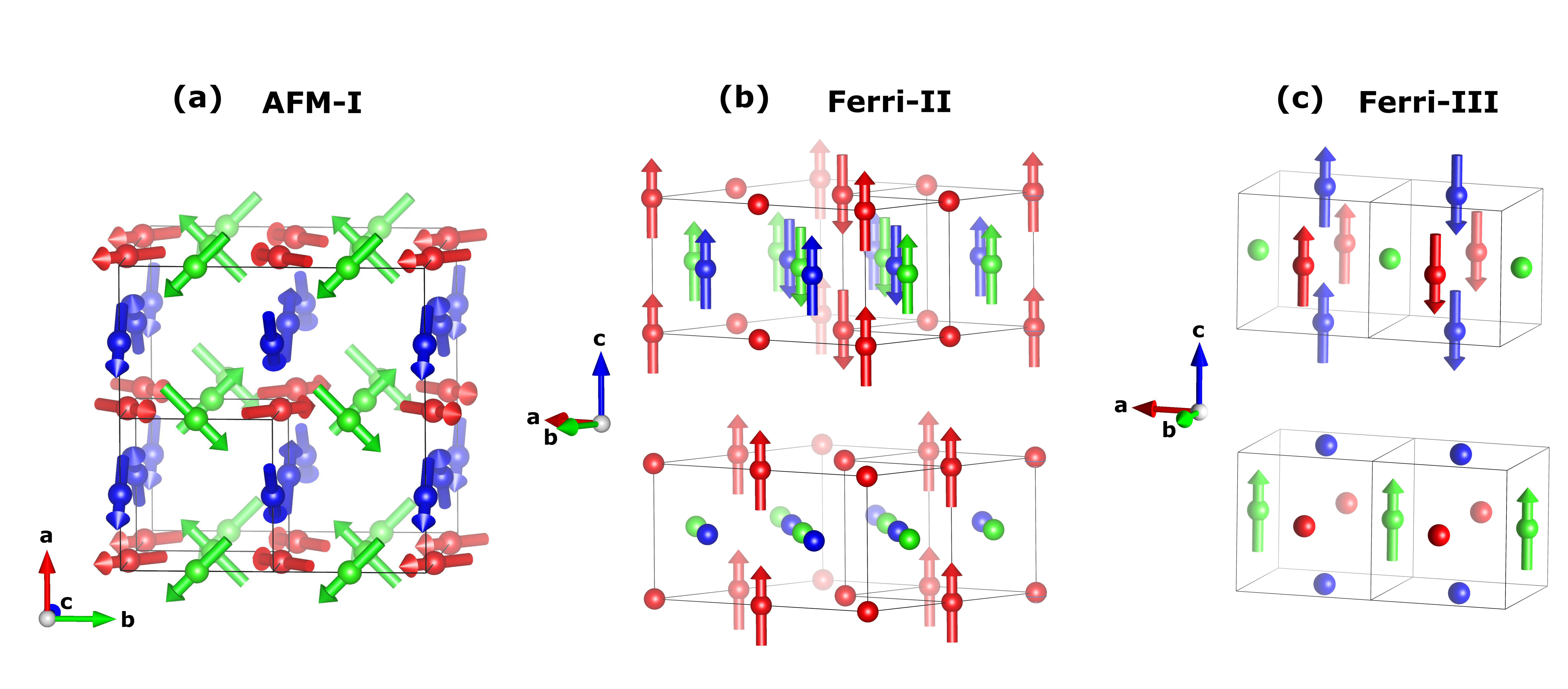}
\caption{
(a) Schematic representation of the AFM-I magnetic structure with the magnetic space group $P_Ia\bar{3}$. (b) Schematic representation of the ferrimagnetic Ferri-II structure with the $P4/mm'm'$ magnetic space group as a superposition of two components, antiferromagnetic (top) and ferromagnetic (bottom). 
(c) Schematic representation of the ferrimagnetic Ferri-III structure with the $Pm'm'm$ magnetic space group as a superposition of two components, antiferromagnetic (top) and ferromagnetic (bottom). 
In all panels three Eu sites associated with the 3c Wyckoff position of the paramagnetic cubic $Pm\bar{3}m$ structure are shown as red, blue, and green spheres, respectively, and the paramagnetic unit cell is shown for clarity alongside the magnetic cells. 
} 
 \label{new_Fig_S7}
\end{figure*}


Eu$_3$PbO crystallizes in the cubic antiperovskite structure with space group $Pm\bar{3}m$ (No.\ 221).
In this crystal structure (inset Fig.~\ref{mFig1}) six Eu atoms form corner-sharing octahedra, which surround the oxygen atoms. 
The Pb atom sits at the corner of the unit cell, surrounded by twelve Eu atoms forming a cuboctahedron. 
We have synthesized Eu$_3$PbO single crystals from stoichiometric amounts of Eu metal together with PbO~\cite{Nuss:dk5032}. Resistivity measurements confirm the expected semimetallic character and preliminary Hall conductivity  measurements indicate hole doping consistent with other inverse perovskites of this series~\cite{2018_prb_Suetsugu,rost_inverse-perovskites_2019}. 
At zero field  Eu$_3$PbO orders antiferromagnetically below the N\'eel temperature $T_{\textrm{N}}$ = 42~K (see Supplementary Note~\ref{supp_note_I} for susceptibility, specific heat and resistivity data) and with increasing applied field undergoes multiple phase transitions. In Fig.~\ref{mFig1}(a) we show magnetization traces for a single crystal of Eu$_3$PbO with the magnetic field oriented along a number of high-symmetry axes. With increasing field the material undergoes several metamagnetic transitions (indicated by arrows) and the spins become fully polarized at the saturation field of approximately 15~T. 
At 20~K the transition into the fully polarized phase is observed at about 12~T, but at 5~K it is just outside the experimentally accessible range of magnetic fields.
Most importantly, the field at which the magnetic phase transitions occur is, to first order, orientation independent. This weak orientation dependence is also of benefit for the powder neutron diffraction measurements in an applied field, with respect to the potential for grain alignment. Our data allow the reconstruction of the overall phase diagram, as shown in Fig.~\ref{mFig1}(b). This is based on powder measurements of specific heat (orange) as well as Quantum Design MPMS (purple) and PPMS (blue) magnetization measurements on a number of samples (see Supplementary Note~\ref{supp_note_I}). 

The phase diagram contains, aside from the high temperature paramagnetic phase, four magnetically ordered states: one antiferromagnetic (AFM-I - labelled I) below $T_{\textrm{N}} = 42$~K and $\mu_0 H $ = 4.6~T, and three ferrimagnetic
within the field range 4.6 -- 15~T (II, III and IV), and a ferromagnetically ordered phase (FM) above approximately 15 T. All phase transitions have a reasonably strong temperature dependence as shown in the phase diagram. The metamagnetic transitions in Eu$_{3}$PbO are reproducible and consistent between multiple batches of powder and single crystal samples, confirming their robust and intrinsic nature.

\subsection{Neutron scattering and magnetic ordering}

\begin{figure*}[btp]
\centering
\includegraphics[width=0.95\textwidth]{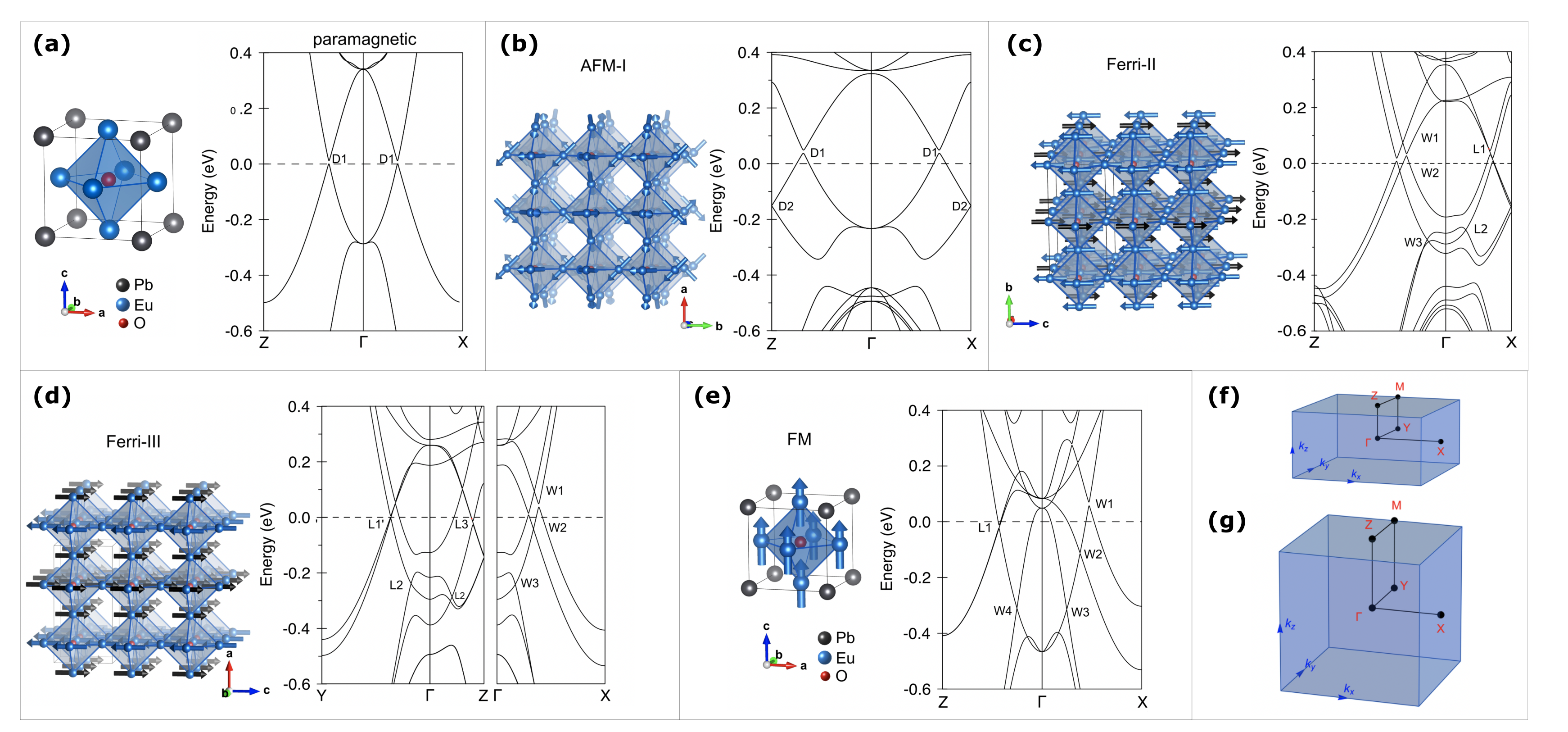}
\caption{\textbf{Electronic band structures, magnetic orders, and Brillouin zones of Eu$_3$PbO.} \label{mFig2}
The Fermi energies are indicated by the dashed lines at $0$~eV. 
The positions of the Dirac points, nodal lines, and Weyl points
are labeled by D, L, and W, respectively, see Table~\ref{mTab1}.
({\bf{a}}) Paramagnetic phase with Brillouin zone shown in ({\bf{f}}). ({\bf{b}}) Antiferromagnetic phase. The paramagnetic zone is folded along $\Gamma-X$, $\Gamma-Y$ and $\Gamma-Z$, giving an 8-fold increase in unit cell size. The cubic symmetry is retained and the bands are Kramers degenerate. ({\bf{c}}) Ferri-II phase. The blue and black arrows indicate the orientations of the magnetic moments on the Eu atoms for clarity. The parent zone is backfolded halfway along $\Gamma-X$ and $\Gamma-Y$. The symmetry is lowered to tetragonal. The bands are now singly degenerate except  at the Weyl points and nodal lines. 
({\bf{d}}) Ferri-III phase with Brillouin zone shown in ({\bf {g}}). The bands are back folded along $\Gamma-Z$ and the symmetry is lowered to orthorhombic. The degeneracy loss is the same as for the Ferri-II phase as described above. 
({\bf{e}}) Fully polarised with field along [110]. This phase has no zone folding but loss of Kramers degeneracy and splitting of Dirac cones into Weyl points. With the field direction along [110] the symmetry is lowered to orthorhombic. Note that while in ({\bf e}) the field is along [110], in all other panels the field is along [100].}
\end{figure*}

In order to determine the  magnetic structure of the four magnetic phases of 
Eu$_3$PbO we performed neutron powder diffraction measurements at the WISH beamline of ISIS, Oxfordshire UK.
Above the N\'eel temperature $T_{\textrm{N}}$ at zero field the refinement yields the cubic space group $Pm\bar{3}m$, with lattice parameter $a= 5.0788(5$)\AA, 
in full agreement with earlier single-crystal x-ray diffraction experiments~\cite{Nuss:dk5032}. 
On cooling below $T_{\textrm{N}}$, we observe clear magnetic Bragg peaks, consistent with propagation vector ${\bf k}= (\frac{1}{2},0,0)$. Given the extremely large number of possibilities of magnetic structures in the current case, the refinement solution underlying our analysis is the highest symmetry one consistent with the magnetic powder neutron diffraction data and the magnetic property measurements. Further details of the magnetic structure solution, along with powder diffraction patterns and fits, are given in Supplementary Note~\ref{supp_note_three}. Solving for the magnetic structure, the magnetic space group $P_Ia\bar{3}$ (No.~205.36, type IV)~\footnote{Here, and in what follows we use the BNS magnetic group type symbols.} gives a satisfactory solution with a non-collinear antiferromagnetic spin alignment, corresponding to the action of the full arms of the propagation vector star, namely ($\frac{1}{2}$,0,0),(0,$\frac{1}{2}$,0) and (0,0,$\frac{1}{2}$) (see Fig.~\ref{mFig2} and Supplementary Note II). This structure can be rationalized with two Eu-Eu interactions: a strong FM interaction through the oxygen and an AFM direct exchange along the octahedral edges
that leads to a 120$^{\circ}$ structure.

Upon increasing the magnetic field, Eu$_3$PbO goes through two metamagnetic transitions at $\mu_0 H = 4.6$~T and $\mu_0 H = 6.1$~T into the two ferrimagnetic phases labelled II and III in Fig.~\ref{mFig1} (Ferri-II and  Ferri-III in Figs.~\ref{new_Fig_S7} and~\ref{mFig2}).
From our neutron measurements, we find that the Ferri-II phase has propagation vectors ${\bf k}=$ ($\frac{1}{2}$,0,0) and (0,0,0). The latter one is consistent with the development of a spontaneous ferromagnetic moment. Fig.~\ref{new_Fig_S7}(b) shows schematically the antiferromagnetic and ferromagnetic components of the ferrimagnetic structure.
The moments are collinear and arranged in (anti)ferromagnetic and ferrimagnetic planes stacked along the \textit{b} direction.
Assuming that the ferromagnetic component is aligned with the field direction, a good quality fit of the magnetic intensities has been achieved in the magnetic space group
$P4/mm'm'$  (No.~123.345, type III)   with the magnetic structure as drawn in Figs.~\ref{new_Fig_S7}(b) and~\ref{mFig2}(c). The magnetic space group derives from the zero field one by the loss of one arm of the star of $\bf{k}$ corresponding to the ferromagnetic direction. 

At fields above $6.1$ T the Ferri-III phase is stabilized, which has a reduced symmetry
with magnetic space group $Pm'm'm$ (No.~47.252, type III) by the loss of another arm of the star of $\bf{k}$, see Figs.~\ref{new_Fig_S7}(c) and~\ref{mFig2}(d).  This magnetic structure consists of ferromagnetic planes stacked along \textit{a} in a mixture of ferromagnetic and antiferromagnetic alignment to result in a net ferrimagnetic, collinear state.
Above the saturation field $\mu_0 H = 15$~T, the last arm of the star is removed and
Eu$_3$PbO becomes ferromagnetic with all spins polarized along the direction of the applied magnetic field.

We note that the identified magnetic structures map well onto the experimental magnetisation data of Fig.~\ref{mFig1}. The magnetisation jumps would be expected to be approximately 1.3$\,\mu_{\textrm{B}}$/Eu$^{2+}$ at each of the 5 T and 8 T transitions, with a remaining moment $\approx$ 5.2$\,\mu_{\textrm{B}}$/Eu$^{2+}$ to reach the fully polarised state. This is in good agreement with our data. We note that there is an approximately linear background to our magnetisation data which may be due to weak spin canting that we cannot determine based on the current polycrystalline sample data (see Supplementary Note \ref{supp_note_three}).

\begin{table}[ht!]
\centering
\begin{ruledtabular}
\begin{tabular}{ccclcc}
phase & position    &  E (eV)  & type & top.~inv. & $\#$ \\ 
\hline 
PM	& {\footnotesize $(0.18,0,0)$}	&  0.017 &  Dirac (D1) & mir.~Chern & 6 \\
\hline 
AFM-I & {\footnotesize $(0.35,0,0)$} & 0.05 &  Dirac (D1)  &  mir.~Chern & 6 \\
AFM-I & {\footnotesize $(\pi,0,0)$} & -0.15 &  Dirac (D2)  &   --
\ & 3 \\
\hline
Ferri-II & {\footnotesize $(0,0,0.17)$} & 0.07 & WP (W1) & Chern & 2\\
Ferri-II & {\footnotesize $( 0, 0,0.18)$} & -0.03 & WP (W2) & Chern & 2\\
Ferri-II & {\footnotesize $(0 ,0, 0.07)$} & -0.27 & WP (W3) & Chern & 2\\
Ferri-II & {\footnotesize $k_x  k_y $-plane} & 0.05 & Line (L1) & Berry & 1  \\
Ferri-II & {\footnotesize $k_x  k_y $-plane} & -0.24 & Line (L2) & Berry & 1\\
\hline
Ferri-III & {\footnotesize $(0.17,0,0)$} & 0.09 & WP (W1) & Chern & 2\\
Ferri-III & {\footnotesize $( 0.16,0,0)$} & -0.04 & WP (W2) & Chern & 2\\
Ferri-III & {\footnotesize $(0.07,0,0)$} & -0.25 & WP (W3) & Chern & 2\\
Ferri-III & {\footnotesize $k_y  k_z $-plane} & -0.003 & Line (L1') & Berry  & 1 \\
Ferri-III & {\footnotesize $k_y  k_z $-plane} & -0.26 & Line (L2) & Berry  & 1 \\
Ferri-III & {\footnotesize $k_y  k_z $-plane} & -0.007 & Line (L3) & Berry  & 2 \\
\hline
FM [110] & {\footnotesize $(0.23, 0.015, 0)$} & 0.06 & WP (W1) & Chern & 4 \\
FM [110]& {\footnotesize $(0.19, -0.001, 0)$} & -0.12 & WP (W2) & Chern & 4 \\
FM [110] &{\footnotesize $(0.12, 0, 0) $} & -0.32 & WP (W3) & Chern & 4 \\
FM [110] & {\footnotesize $( 0.003,  0.003 ,0.13)$} & -0.31 & WP (W4) & Chern & 4 \\
FM [110] & {\footnotesize [110]-plane} & -0.02 & Line (L1) & Berry  & 1 \\
\end{tabular}
\end{ruledtabular}
\caption{
\textbf{Types of topological band crossings.}
This table lists the positions and energies of the topological band crossings in the first Brillouin zone (BZ)
for the paramagnetic phase (PM), the antiferromagnetic phase (AFM), the   ferrimagnetic phases (Ferri-II and Ferri-III),
and the ferromagnetic phase with magnetization in [110] direction (FM [110]).
 The positions of the band crossings ${\bf k} = ( k_x, k_y, k_z)$ are given in 
units of $2 \pi/ a_{i}$, where $a_i$ denotes the lattice constant of the respective real space direction. All energies are given in eV relative to the Fermi energy. 
The type of band crossing is indicated in the fourth column, while the fifth column 
states the topological invariant that protects the crossings.
The last column gives the multiplicity of the crossings, i.e., the number of symmetry related crossings
at the same energy.
\label{table_band_crossings} 
\label{mTab1}
}
\end{table}

\subsection{Topological band structure}

\begin{figure*}[hbt]
\centering
\includegraphics[width=0.93\textwidth]{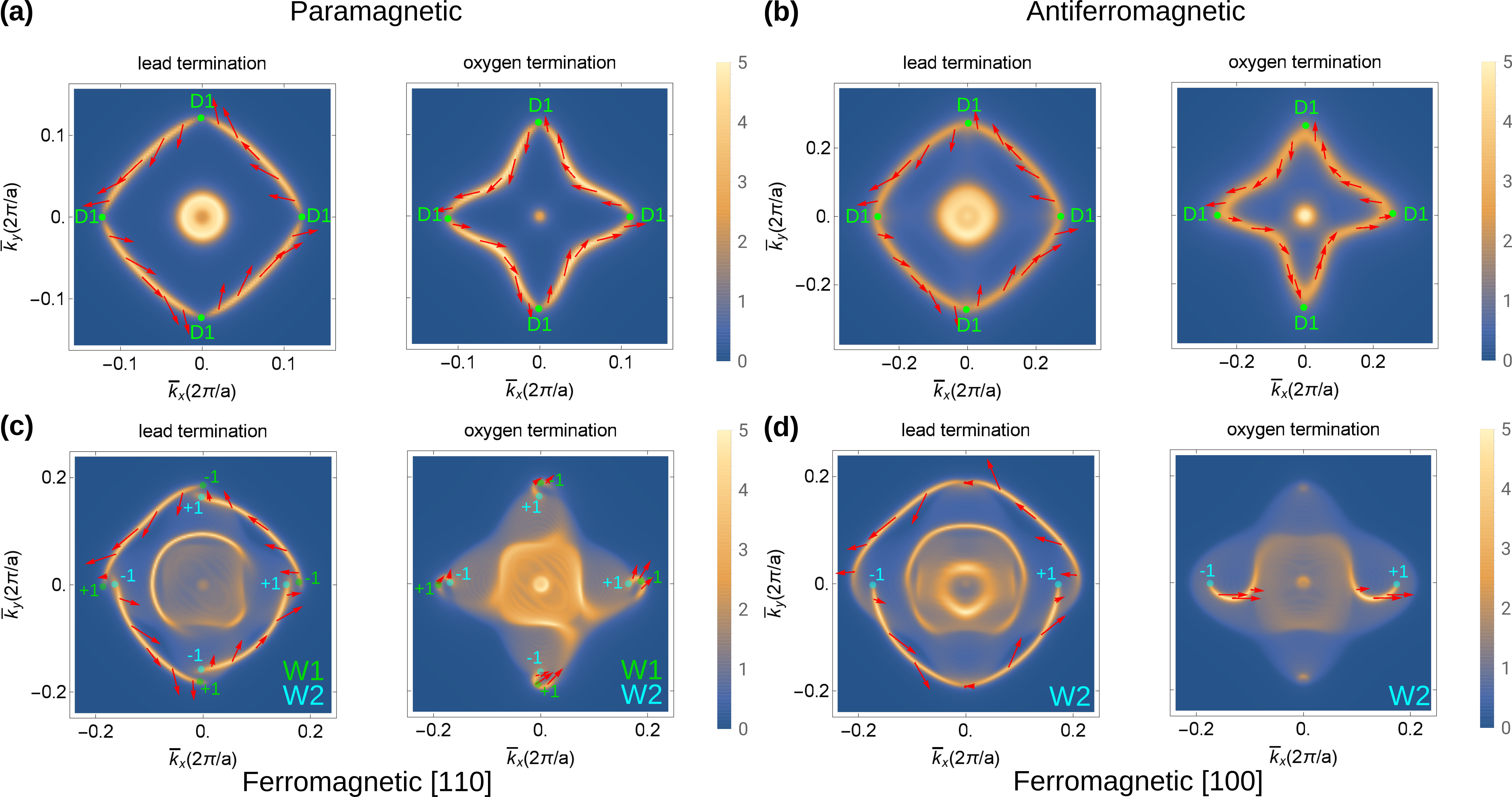}
\caption{\textbf{Surface states of Eu$_3$PbO and their spin polarization.} \label{mFig3}
Calculated surface density of states (SDOS) and spin polarization for a  [001] slab of Eu$_3$PbO  with oxygen and lead termination.
The color code represents the SDOS on a linear scale for the five outermost layers, while the in-plane spin polarization is indicated by the red arrows. 
({\bf a}),({\bf b}) show the SDOS for the paramagnetic and antiferromagnetic phases at the energy of the upper Dirac points $E=0.017$~eV and $E=0.05$~eV, respectively (cf.~Table~\ref{table_band_crossings}). The position of the Dirac points is marked by the label D1.
({\bf c}), ({\bf d}) display the SDOS for the ferromagnetic phase with magnetization in [110] and [100] directions at the  energy of  the Weyl points W1 
and W2,  respectively. The chiralities and positions of the Weyl points W1 and W2 are indicated in all panels by the green and blue dots and numbers.
}
\end{figure*}

In the paramagnetic phase the band structure of Eu$_3$PbO displays six three-dimensional gapped Dirac cones at finite momentum along
the $\Gamma$--$X$, $\Gamma$--$Y$, and $\Gamma$--$Z$ directions, see Fig.~\ref{mFig2}(a) and Table~\ref{mTab1}.
This is similar to the nonmagnetic antiperovskite Ca$_3$PbO~\cite{kariyadoJPSJ11,kariyadoJPSJ12,chiu_PRB_17}, where a linear Dirac-like dispersion
has recently been observed using soft x-ray ARPES measurements~\cite{obata_hosono_Ca3PbO_arpes_PRB_17}.
Indeed, just as in Ca$_3$PbO~\cite{hsieh_fu_PRB_14,chiu_PRB_17}, the paramagnetic state of Eu$_3$PbO is a crystalline topological insulator 
characterized by two independent mirror Chern numbers~\cite{chiu_PRB_17}.
This nontrivial topology arises due to a band inversion  of the Eu-$d$ and Pb-$p$ orbitals near the $\Gamma$ point and is protected
by the mirror symmetries of the cubic space group $Pm\bar{3}m$. 
At the surface, the nontrivial band topology manifests itself by the appearance of two two-dimensional Dirac cone surface states, like in Ca$_3$PbO~\cite{kariyadoJPSJ12,chiu_PRB_17}, see Fig.~\ref{mFig3}(a).

While the PM phase is very similar to the non-magnetic antiperovskites, the magnetic phases realize a number of novel topological states.
Indeed, it is expected that the different magnetic orders have significant effect on the band structure of Eu$_3$PbO, leading to different topological states.
Let us now discuss the topological band structures of the three different magnetic phases in detail. 

\subsubsection{Antiferromagnetic phase}

Similar to the PM phase, all bands in the AFM-I phase are two-fold degenerate, even though time-reversal symmetry $\mathcal{T}$ is broken.
This is because magnetic space group $P_Ia\bar{3}$ (No.~205.36) contains a symmetry element $\widetilde{\mathcal{T}}$ that combines time-reversal 
with a half translation along [111]  (equivalent to \{1$^{\prime}\mid\,\frac{1}{2},\frac{1}{2},\frac{1}{2}\}$), leading together with inversion $P$ to  Kramer's degeneracies at every ${\bf k}$ point.
Since the AFM-I phase has an eight times larger unit cell compared
to the PM phase,  the electronic bands are backfolded in all three reciprocal directions, as seen by comparing~Fig.~\ref{mFig2}(a) with Fig.~\ref{mFig2}(b).
Backfolding leaves the six gapped Dirac points D1 of the PM phase mostly unchanged, although they are moved to a slightly higher energy (cf.~Table~\ref{mTab1}).
Along the $\Gamma$--$X$, $\Gamma$--$Y$ and $\Gamma$--$Z$ directions the backfolded bands hybridize and a band gap of about $100$~meV opens up. 
At the time-reversal invariant momenta $X$, $Y$, and $Z$, however, 
hybridization is strongly suppressed by symmetry, giving rise to three (nearly) gapless Dirac points
at the energy $E \simeq -0.15~\mathrm{eV}$ [D2 in Fig.~\ref{mFig2}(b)]
\footnote{There exists a very small, but finite hybridization between the Pb-$p$ and Eu-$d$ orbitals at the $X$, $Y$, and $Z$ points, 
which leads to a small gap of less than $1$~meV, see Supplementary Note~\ref{TB_model}.}.
Notably, the dispersion at these Dirac crossings is linear in one direction,
but quadratic in the other two. 

We conclude that the AFM-I phase of Eu$_3$PbO contains both gapped and gapless Dirac points close to the Fermi energy, within an experimentally accessible energy range. 
Remarkably, this is only the second known example of a Dirac state in an antiferromagnet, the other one being CuMnAs~\cite{tang_zhang_nat_phys_2016}.

\subsubsection{Ferrimagnetic phases}

In the ferrimagnetic phases Ferri-II and Ferri-III the symmetry is lowered, which results in a lifting of the Kramer's degeneracy of the bands.
As a consequence, all bands are in general singly degenerate, but can form doubly degenerate Weyl points when they cross. 
In Table~\ref{mTab1} we list the Weyl points (WP) that are close to the Fermi energy, see Fig.~\ref{mFig2}(c) and~\ref{mFig2}(d).
We observe that the multiplicity of all the Weyl points is only two, i.e., due to the low symmetry there are only two Weyl points at the same energy.
Such a pair of Weyl points, which are related by inversion, represents the simplest kind of Weyl band crossings. 
The low-energy physics near these Weyl points is described by magnetic Weyl fermions, which exhibit a number
of exotic phenomena, e.g., topological (magneto)transport properties due to the chiral anomaly~\cite{armitage_mele_vishwanath_review,burkov_review_2018}.
By doping Eu$_3$PbO with, e.g., Eu deficiencies, the Fermi level could be tuned to these pairs of Weyl points, which would allow 
to measure the topological transport signatures in a clear way. 

The Ferri-II and Ferri-III phases exhibit in addition to the Weyl points also line crossings, 
where  two bands intersect along a one-dimensional line in the BZ.  In the Ferri-II phase these line crossings  
appear in the $k_x k_y$-plane 
and are mapped onto themselves under inversion, resulting in multiplicity one.
They are protected by the mirror symmetry $z \to -z$ and a $\pi$-Berry phase, which is expected to lead to drumhead surface states~\cite{nodal_line_Yang}, whose shapes depends on the chosen termination~\cite{Rhim2017BulkBoundaryZak}.
In the Ferri-III phase the line crossings occur  in the $k_y k_z$-plane and are protected by the mirror symmetry $x \to - x$, 
since here the magnetic moments point along the $x$ direction, rather than the $z$ direction.
Contrary to the Ferri-II phase, the Ferri-III phase also exhibits a pair of nodal lines with multiplicity two (L3 in Table~\ref{mTab1}).
The two nodal lines of this pair are mapped onto each other by inversion symmetry.

\subsubsection{Ferromagnetic phase}

In the FM phase all magnetic moments are colinearly aligned along the direction of the applied magnetic field.  
As a consequence, the Kramers degeneracy of the bands in the PM and AFM-I phases is lifted and the gapped Dirac points
are split up into a collection of Weyl points. 
For concreteness, we consider here a field applied along the [110] direction, which lowers
the symmetry to $Cmm^{\prime}m^{\prime}$  (No.~65.486) and leads
to an interesting set of Weyl points and nodal lines (other field directions are discussed Supplementary Note~\ref{supp_note_one}). 
In this case, the bands near the Fermi energy form  four quartets of Weyl points and one nodal line, 
see Fig.~\ref{mFig2}(e) and Table~\ref{mTab1}.
The four Weyl points of the quartets W1, W2, and W3 are located within the $k_z = 0$ plane
and are symmetry related by  inversion and mirror symmetry $k_x \leftrightarrow k_y$. 
The four Weyl points of the quartet W4 lie within the [$\bar{1}$10] plane and
are symmetry related by inversion and two-fold rotation along the $z$-axis combined with time-reversal. 
We remark that opposite chirality Weyl points in the quartets W1, W2, and W3 show significant separation in momentum
space, which results in large arc surface states (cf.~Fig.~\ref{mFig3}) and, moreover, enhances
the topological transport signatures. 
Opposite chirality Weyl points in the quartet W4, on the other hand, are close together in ${\bf k}$ space, separated
by only $\delta k = 0.008 \times 2\pi / a $, see  Table~\ref{mTab1}. 
Besides these Weyl points, the FM phase with [110] magnetization exhibits
also a line node in the plane perpendicular to the [110] direction, which is protected by mirror symmetry.

\subsection{Surface states}

The surface states in the PM phase of Eu$_3$PbO,
which are shown in Fig.~\ref{mFig3}(a),
are  similar to the nonmagnetic antiperovskites.
Like in Ca$_3$PbO~\cite{chiu_PRB_17,kariyadoJPSJ12}, we observe two-dimensional
Dirac cone surface states, both for the lead and oxygen terminations~\footnote{
Lead (oxygen) termination refers to surfaces that contain besides europium only lead (oxygen) atoms, cf.~Ref.~\cite{chiu_PRB_17}.}.
These Dirac cone surface states appear by the bulk-boundary correspondence, as a consequence of the nonzero mirror Chern numbers that characterize 
the nontrivial bulk topology. 
Since the surface states are singly degenerate, they exhibit a nontrivial spin texture, as indicated by the red arrows.

In the magnetically ordered phases of Eu$_3$PbO,  the ordered Eu moments cause large changes not only in the bulk bands but also in the surface states. 
To exemplify this, we focus on the surface states of the AFM-I and FM phases.

\subsubsection{Antiferromagnetic phase}

In Fig.~\ref{mFig3}(b) we present the surface states of the AFM-I phase with [001] termination. We find 
that the surface states are qualitatively similar to the PM phase. For both the lead and the oxygen termination
there appear Dirac cone surface states with a nontrivial spin polarization.
Because of the backfolding of the BZ, these surface states now cover twice as much area as in Fig.~\ref{mFig3}(a).
As in the PM phase, the Dirac cone surface 
states of the AFM-I phase are guaranteed to exist due to the bulk-boundary correspondence, 
which relates them to the  mirror Chern number of the bulk bands.

\subsubsection{Ferromagnetic phase}

In the FM phase the nontrivial topology of the Weyl points leads to the appearance of arc surface states,
whose stability is guaranteed by a nonzero Chern number.
The surface states on the [001] surface of the FM phase with magnetization direction [110]   
are shown in Fig.~\ref{mFig3}(c). For the lead termination we observe
four large arc states that connect opposite chirality Weyl points of the quartets W1 and W2.
Since these arc states extend over nearly half the surface BZ, they should be readily observable
via quasi-particle interference in Fourier-transform scanning tunneling spectroscopy~\cite{moessner_QPI_Weyl}.
For the oxygen termination, on the other hand, the arc states are much shorter,
connecting opposite chirality Weyl points that are located next to each other.
Regarding the spin polarization of the arc states, we find that for the lead termination
the polarization is similar to the one of the Dirac states in the PM and AFM-I phases. 
For the oxygen termination, however, the spin polarization points predominantly along
the direction of the Eu moments. This is because for the oxygen termination, there
are no Pb atoms and twice as many Eu atoms on the surface compared to the lead termination. Hence, the surface states 
on the oxygen termination have mostly Eu-$d$ orbital character, 
while for the lead termination they have Pb-$p$ character.
For this reason the surface states on the oxygen termination are polarized more
strongly by the Eu moments than on the lead termination. 

So far we have assumed that the Eu moments point along the [110] direction. However, in a single crystal it is possible
to adjust the magnetization direction  by aligning the spins using the external field. 
This allows us to tune the electronic structure and band topology of Eu$_3$PbO.
To demonstrate this, let us consider the FM phase with the Eu moments oriented along
the [100] axis. With this magnetization direction, the Weyl points W2 and W3 are slightly shifted in energy,
while the W1 points are entirely absent (Supplementary Note~\ref{supp_note_one}).
Correspondingly, the arc surface states show   different connectivities, see Fig.~\ref{mFig3}(d).
I.e., on the [100] surface there appears a single arc state that connects the two Weyl points W2,
both for the lead and oxygen terminations. (For the lead termination there is a secondary surface state
feature, which however has no topological origin.)

\begin{figure}[t!]
\centering
\includegraphics[width=0.43\textwidth]{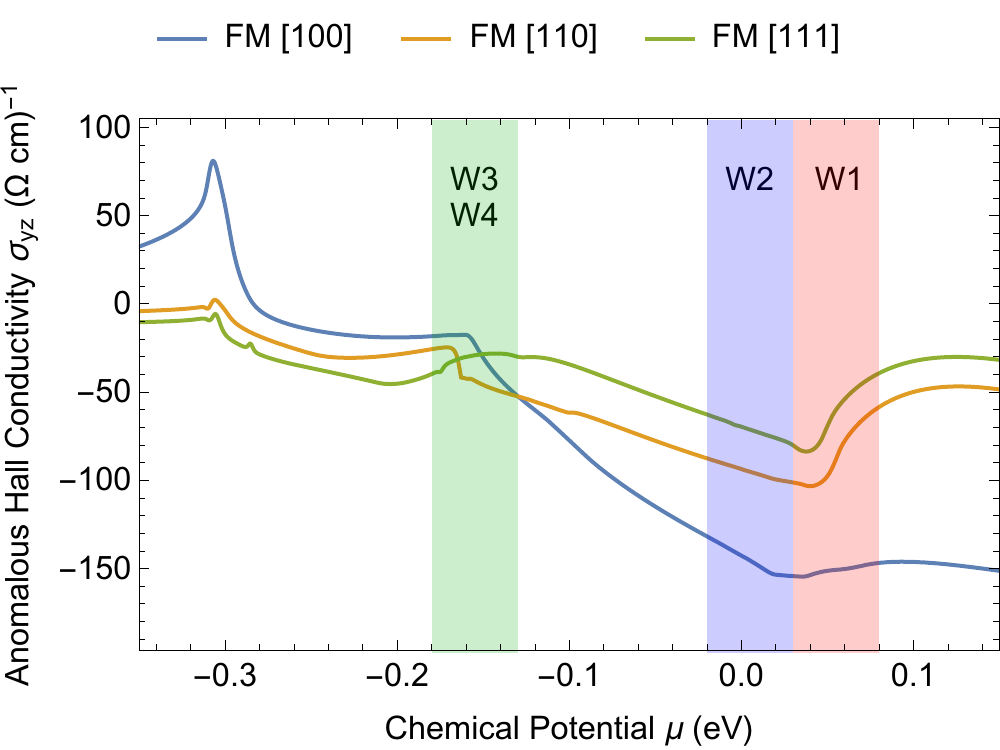}
\caption{ \label{fig_berry_curvature} \label{mFig4} 
\textbf{Intrinsic anomalous Hall conductivity.}
Calculation of the intrinsic anomalous Hall conductivity $\sigma_{yz}$  in the ferromagnetic phase as a function of chemical potential $\mu$, for different
magnetization directions at a temperature of 10~K.
Shaded regions highlight broad features in  $\sigma_{yz}$ that are attributed
to different Weyl points.  
}
\end{figure}

\subsection{Anomalous Hall conductivity}

The nontrivial band topology of Eu$_3$PbO manifests itself not only in 
the surface states but also in anomalous transport characteristics, such as the anomalous Hall effect,
the anomalous Nernst effect~\cite{ikhlas_nakatsuji_Mn3Sn_Nernst_nat_phys_17}, or the  circular photogalvanic effect~\cite{cpge_moore_nat_commun_17}.
For instance, the intrinsic anomalous Hall conductivity (AHC) $\sigma_{i j}$
is directly proportional to the momentum integrated Berry curvature. 
The AHC shows local extrema when the
chemical potential is tuned to the Weyl point energies. 
Moreover, the dependence of the band topology on the magnetic phase and the magnetization direction
is expected to reflect itself in the AHC.

To exemplify this, we calculate the intrinsic AHC $\sigma_{yz}$ for the FM phase with different magnetization directions 
as a function of chemical potential $\mu$ (Fig.~\ref{mFig4}). 
We observe that the overall magnitude as well as the position of the local extrema changes
with magnetization direction. For small hole doping
at $\mu \simeq -0.05$~eV,
the AHC in the FM phase with [110] and  [111] magnetization is about 50~$(\Omega \textrm{cm})^{-1}$,
while in the FM phases with [100]  magnetization the AHC is three times larger.
This is quite sizable and comparable to the values of MnSi~\cite{manyala_fisk_MnSi_nature_04} and Mn$_3$Sn~\cite{higo_magnetic_weyl_anom_hall_nat_15}.
The broad features in $\sigma_{yz}$ at $\mu \simeq +0.05$ and $\mu \simeq -0.15$  originate from the Weyl 
points W1/W2 and W3/W4, respectively (Supplementary Note~\ref{supp_note_two}). Interestingly,
at $\mu \simeq +0.05$ W1 gives a positive contribution, while W2 contributes negatively, because these
two sets of Weyl points have opposite chiralities. Due to this cancellation, the AHC   for the [110] and [111] magnetizations
is about twice smaller than for the [100] magnetization, for which the Weyl points W1 do not exist.

\section{Summary and discussion}

Combining magnetization measurements with neutron diffraction and electronic structure calculations, we have studied the 
interplay between band topology and magnetism in the antiperovskite Eu$_3$PbO. 
We have discovered four different magnetic phases 
and identified their magnetic structures as a function of magnetic field. 
For each of these phases we have determined the band topology,
thereby uncovering a rich variety of Weyl points, Dirac points, and nodal lines
close to the Fermi energy. 
By the bulk-boundary correspondence,
this nontrivial topology of the bulk bands leads 
to various types of surface states, e.g., Dirac cone, Fermi arc, and drumhead surface states,
within an easily accessible magnetic phase diagram in a single compound. 
Moreover, the bulk topology gives rise to
unusual transport phenomena, such as anomalous Hall currents.
We have calculated the anomalous Hall conductivity of Eu$_3$PbO in the ferromagnetic phase
and shown that it displays clear fingerprints of the Weyl points. 
At the metamagnetic transitions the anomalous Hall current exhibits sharp singularities, due
to the rearrangement of the magnetic spin texture. 
Thus, the different (noncollinear) magnetic orders in Eu$_3$PbO offer the unique opportunity
to explore the sensitivity of the anomalous Hall current on the details of the magnetic structure. 

The four magnetic phases of Eu$_3$PbO with their different band topologies  can be easily accessed 
and manipulated with an external field. This allows to tune the electronic structure 
and drive it through topological phase transitions. For example, the Dirac points
of the antiferromagnetic phase can be split into Weyl points by crossing the phase boundary into
to ferrimagnetic phase. Furthermore, the Weyl points and line nodes of the ferri- and ferromagnetic phases
can be pair-annihilated or moved in energy and momentum by adjusting the magnetization direction.
This, in turn, modifies the Berry curvature of the bands, and hence the anomalous Hall conductivity. 
Thus, Eu$_3$PbO offers a rich playground to study the interdependence among magnetism,  topology of the electronic bands, and
anomalous transport properties.
This makes Eu$_3$PbO a potential candidate for new device
applications that rely on magnetic-field induced switching of  topological currents, especially in light of the advances in the thin film growth of the related compound Sr$_3$PbO \cite{smal_thin_film_Sr3PbO_APL_mat_16}.

In closing, we discuss several possible directions for future experimental and theoretical studies.  
First, the bulk and surface Dirac cones of the paramagnetic and antiferromagnetic phases could 
be measured by ARPES, since these phases have no net magnetic moment.
The bulk Weyl cones and surface arcs of the ferri- and ferromagnetic phases, on the other hand,
could be observable in Fourier-transform scanning tunneling spectroscopy.
Second, Hall resistivity and magnetoconductance measurements on single crystals are of high interest,
as these would reveal transport signatures of the Weyl points and nodal lines.
Third, Nernst effect and magnetothermal transport measurements   
 could provide a direct measure of the Berry curvature~\cite{ikhlas_nakatsuji_Mn3Sn_Nernst_nat_phys_17}.
Furthermore, they  could reveal possible violations of the  Wiedemann-Franz law, 
due to the chiral anomaly of the Weyl points~\cite{sharma_goswami_tewari_PRB_16}.
We note that single crystals of Eu$_3$PbO are naturally hole doped,
such that the Dirac  and Weyl points D2 and W2 should be readily accessible
in transport and ARPES measurements.
On the theoretical side, it would be interesting
to investigate the RKKY interactions among the Eu moments 
and to study effects of magnetic and Coulomb interactions and disorder
on the band topology.  
We hope that our work will inspire future research along these lines.


\vspace{0.5cm}

 
\section{Methods}

\subsection{Electronic structure calculations and tight-binding model}

The electronic structure of Eu$_3$PbO was computed by a   relativistic linear muffin-tin orbital calculation~\cite{andersenPRB75,book:AHY04}
 using the in-house PY LMTO computer code as described in Ref.~\cite{book:AHY04}. The code is available on demand.
 As an input for the DFT calculation we used the experimental crystal structure of Ref.~\cite{Nuss:dk5032}.
The DFT calculations show that the bands near the Fermi energy originate mostly from 
Pb-$p$ and Eu-$d$ orbitals.
Guided by these observations, we use the Pb-$p$ and Eu-$d$ orbitals as a basis set to derive
a nine-band tight-binding model. 
With this tight-binding model we have computed the surface states, the Berry curvature, and the topological invariants.
The details of the tight-binding model are presented in Supplementary Note~\ref{supp_note_one}.

\subsection{Topological invariants, surface states, and anomalous Hall conductivity}

The stability of the Dirac, Weyl, and nodal-line band crossings is guaranteed by nonzero topological invariants.
We have numerically computed these topological invariants using 
the  tight-binding model,
see Supplementary Note~\ref{supp_note_two}. 
By the bulk-boundary correspondence a nonzero value of the topological invariant
leads to protected surface states. Using the tight-binding model we have
computed the density of states and spin polarization of these surface states, which are presented in Fig.~\ref{mFig3}. 
The anomalous Hall conductivity is obtained from the momentum 
integral of the Berry curvature, see Supplementary Note~\ref{supp_note_two}.

\subsection{Crystal growth and sample characterization}

Crystals of Eu$_3$PbO were grown in a sealed Ta ampoule as reported previously in ~\cite{Nuss:dk5032}. Since Eu$_3$PbO is extremely air sensitive, all experiments were prepared and sealed under Ar atmosphere. Single crystals for magnetization measurements were covered by a thin layer of Apiezon N grease to protect it from air during transfer. All magnetization measurements were performed both in a Quantum Design SQUID as well as with the VSM option in a Quantum Design PPMS system.

\subsection{Neutron scattering measurements}

Neutron scattering measurements were performed at the WISH beamline of ISIS, Harwell Oxford. Polycrystalline Eu$_3$PbO was used for all of the neutron diffraction measurements. Isotope enrichment was not possible due to the necessity of using EuO as a reagent, rather than Eu$_{2}$O$_{3}$. In zero field a loose powder was contained in a cylindrical 1 mm diameter suprasil capillary (sealed under 200 mbar He to provide exchange gas and prevent sample decomposition) which was then held in a vanadium can for the measurement. A pelletised sample was used for the applied field measurements to prevent grain alignment in applied field, this was contained in a flattened suprasil ampoule to prevent movement and again was sealed under He atmosphere. Data were collected at 50 K and 1.5 K for the loose powder sample and 50 K at 0 T, 1.5 K at 0 T followed by 1.5 K at 5 T and 1.5 K at 8 T without intermediate warming. For the details on the magnetic structure determination see Supplementary Note~\ref{supp_note_three}.

\bibliographystyle{naturemag}

\bibliography{Eu3PbO}

\vspace{0.5 cm}
 
\noindent
\textbf{Acknowledgements} \\
 We gratefully acknowledge useful discussions with A.~Bangura, A. Leonhardt, and H. Nakamura.  
 We thank the Science and Technology Facilities Council for beamtime under RB1700035. 
 We wish to thank R.~Kremer and  E.~Br\"ucher for physical property measurements. A. W. R. and A. S. G. were supported by the Engineering and Physical Sciences Research Council (grant numbers EP/P024564/1 and EP/T011130/1 respectively). This work has been supported in part by the Alexander von Humboldt Foundation.

\clearpage
\newpage


 \setcounter{figure}{0}
\makeatletter 
\renewcommand{\thefigure}{S\@arabic\c@figure} 

\setcounter{equation}{0}
\makeatletter 
\renewcommand{\theequation}{S\@arabic\c@equation} 

 \setcounter{table}{0}
\makeatletter 
\renewcommand{\thetable}{S\@Roman\c@table} 

\makeatother

\setcounter{section}{0}
\refstepcounter{section}
\onecolumngrid
\section*{Supplementary Note I: Specific Heat and Susceptibility Measurements} \label{supp_note_I}

\begin{figure}[h]
\begin{center}
\includegraphics[width=0.99\textwidth]{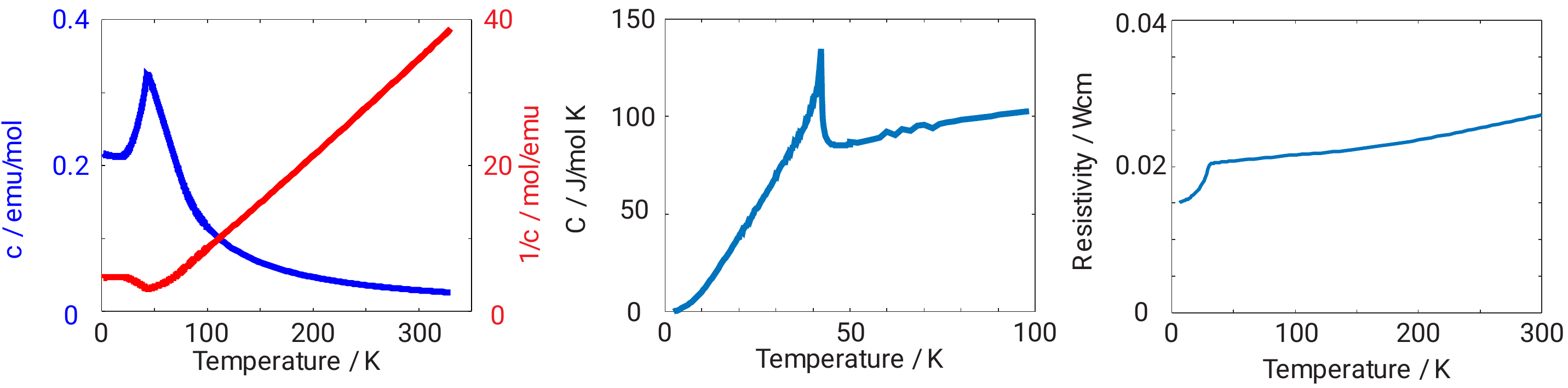}
\end{center}
\caption{Magnetic susceptibility (left), specific heat (middle), and resistivity (right) of Eu$_3$PbO. For more details see text. }
  \label{fig:figsup1}
\end{figure}
Here we briefly present further physical property measurements underlying the phase diagram presented in the main text. In Fig. \ref{fig:figsup1} we present magnetic susceptibility (top), specific heat (middle), and resistivity (bottom) data of Eu$_3$PbO. 

The magnetic susceptibility was measured on a powder sample in a field of 100 Oe and shows a clear transition into the antiferromagnetic phase at app. 40 K. Above $T_N$ a clear Curie-Weiss behaviour is observed corresponding to a magnetic moment of 7.8 $\mu_B$ consistent with the expected moment for Eu$^{2+}$.

Specific heat was measured in zero field on a pressed pellet in a Quantum Design PPMS system. The transition into the antiferromagnetic state is clearly observable at $T_N$.

Finally we present resistivity measurements again on a pressed pellet. While low temperature transport is dominated by grain boundaries the semi-metallic behaviour at higher temperatures as well as the transition into the antiferromagnetic state is readily discernible. Especially the latter indicates a strong coupling of the magnetic order with electronic transport.

\clearpage
\newpage

\refstepcounter{section}
\section*{Supplementary Note II: Determination of the magnetic structures} \label{supp_note_three}

Polycrystalline Eu$_3$PbO was used for all of the neutron diffraction measurements. In zero field a loose powder was contained in a cylindrical 1 mm diameter suprasil capillary (sealed under 200 mbar He to provide exchange gas and prevent sample decomposition) which was then held in a vanadium can for the measurement. A pelletised sample was used for the applied field measurements to prevent grain alignment. This was contained in a flattened suprasil ampoule to prevent movement under applied field and again was sealed under He atmosphere. Data were collected at 50 K and 1.5 K for the loose powder sample and 50 K at 0 T, 1.5 K at 0 T followed by 1.5 K at 5 T and 1.5 K at 8 T without intermediate warming. 
\vspace{0.4cm}

\subsection{Crystal Structure of Eu\textsubscript{3}PbO}

At 50 K in zero field a good structural refinement could be obtained in $Pm\bar{3}m$, in good agreement with previous powder and single crystal x-ray diffraction studies, with $a=5.0788(5)$\AA. The setting with Eu on Wyckoff site $3c$ (0,1/2,1/2), Pb on site $1a$ (0,0,0) and O on site $1b$ (1/2,1/2,1/2) was used throughout.

\subsection{Zero field magnetic structure}

\begin{figure}[h]
\begin{center}
\includegraphics[width=0.55\textwidth]{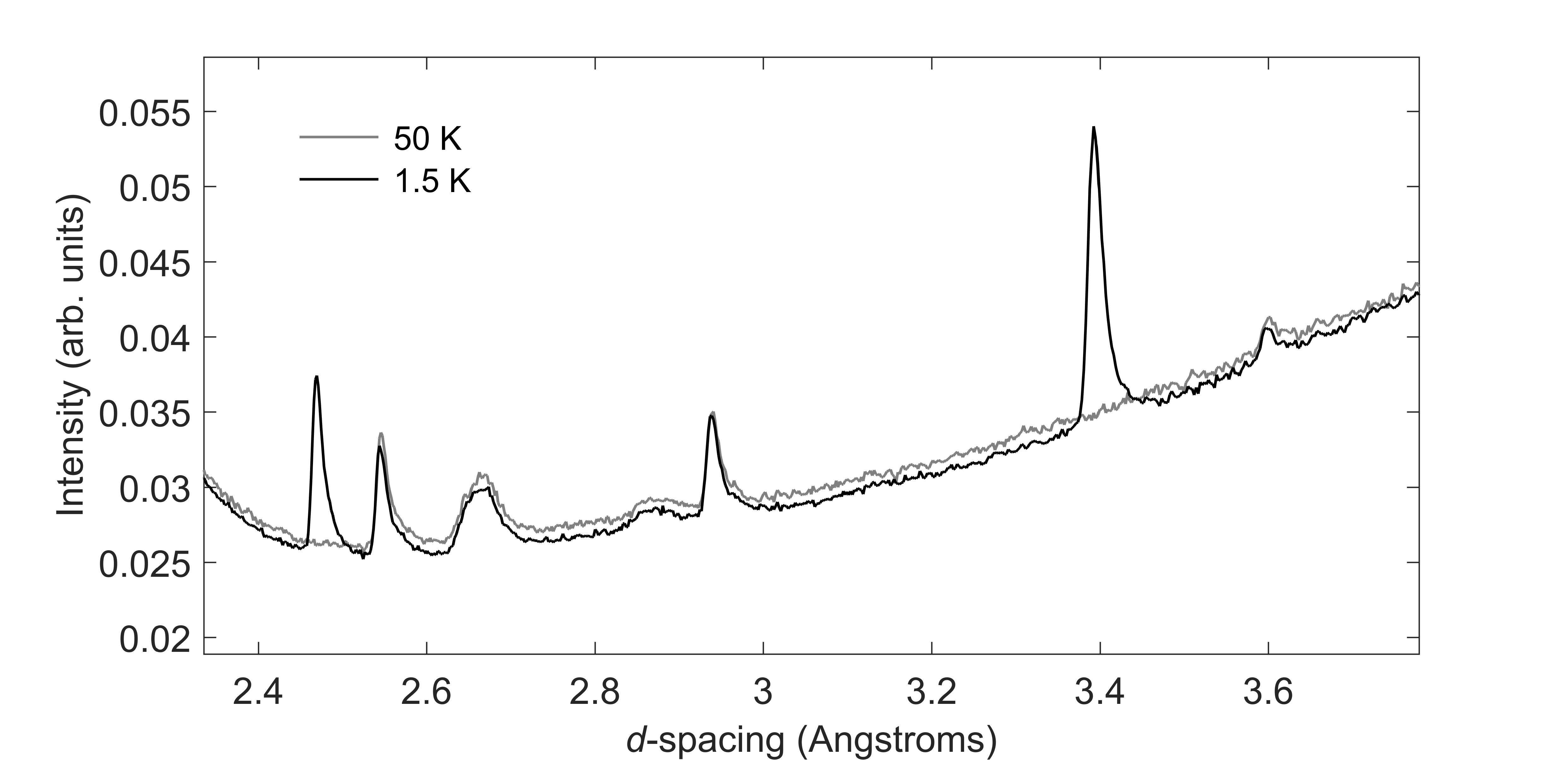}
\end{center}
\caption{The paramagnetic (50 K) and antiferromagnetic (1.5~K) phase data from banks 3 \& 8 of WISH. The substantial background is from the quartz ampoule used to contain the sample.}
  \label{fig:figs1}
\end{figure}

In the 1.5 K data magnetic Bragg peaks were observed (see Fig.~\ref{fig:figs1}), consistent with a propagation vector of $\mathbf{k}=(1/2,0,0)$. The magnetic structure solution began with the assumption (supported by magnetometry) that a finite ordered moment must be present on all Eu sites. The maximal magnetic subgroups of \pmp{} for propagation vector $\mathbf{k} = (1/2,0,0)$, considering a single arm only, and a magnetic ion on the $3d$ site were investigated using the MAXMAGN tool of the Bilbao Crystallographic Server~\cite{Mato_etal_AnnRevMatRes_MAXMAGN_2015}, but none satisfied the above condition. 

Following this, the tool k\_Subgroupsmag~\cite{Perez_Mato_kSubgroupsmag_2016} was used to determine the subgroups of \pmp{} which fulfill this condition. These subgroups were investigated in order of decreasing symmetry by examination of systematic absences followed by Rietveld refinement using FullProf and Jana2006 of potential candidates \cite{1993_55_RODRIGUEZCARVAJAL_fullprof, JANA2006}. 

Systematic absences violations excluded $P_{c}\bar{4}c2$, $P_{c}4cc$, and $P_{c}cc2$. Rietveld refinements were performed using magnetic-only datasets created by subtracting the 50 K paramagnetic data from the 1.5 K data. The scale and absorption factors were fixed by a refinement of the nuclear phase with the 50 K dataset. $A_{a}ma2$ (BNS 40.208) with transformation matrix (2$c$, -$a+b$, $a+b$), origin shift (0,0,1/2) was the only direct subgroup of \pmp{} in the tree under consideration found to give a satisfactory refinement and therefore no further symmetry lowering was investigated. The results of the refinement can be found in Table \ref{tab:tab2}. The moment obtained for Eu1 is slightly in excess of the expected ordered moment. In addition, the moments on the two Eu sites are perpendicular which is not easily rationalized. Since the different moment sizes as well as orientation are not reconcilable with physical property measurements we can exclude this single-$k$ solution.

\begin{table}[!t]
  \centering
   \caption{The results of the Rietveld refinement of the $1.5\,$K 0 T Eu$_3$PbO data in magnetic space group $A_{a}ma2$, in the standard setting where a=$\sqrt2a_p$, b=$\sqrt2a_p$, c=$2a_p$, and $a_p$ is the parent lattice constant $a_p=5.0763(13)$\AA. }
\vspace{0.3cm}
\begin{ruledtabular}
  \begin{tabular}{ccccccc}
   Site & x & y & z & $m_{x}$ & $m_{y}$ & $m_{z}$  \\ 
    &  &  &  & ($\mu_{B}$/Eu$^{2+}$)& ($\mu_{B}$/Eu$^{2+}$) & ($\mu_{B}$/Eu$^{2+}$)\\
\hline
Eu1 & 0.75 & 0.25 & 0.25 & 7.49(8) $\pm$0.08 & 0 & 0  \\  
Eu2 & 0 & 0 & 0.0 & 0 & 0 & 2.57 $\pm$0.14 \\ 
  \end{tabular} 
  \end{ruledtabular}
  \label{tab:tab2}
\end{table}

However, the data are consistent with a multi-$k$ structure with the use of the full arm of the star - $\mathbf{k}_{1}=(1/2,0,0)$, $\mathbf{k}_{2}=(0,1/2,0)$, $\mathbf{k}_{3}=(0,0,1/2)$. There are a large number of possible structures consistent with these vectors, with differing moment directions relative to the lattice. However, in the cubic metric this cannot be determined from powder data. Therefore, the highest symmetry option consistent with the data, $P_{I}a\bar{3}$ with a single Eu site, was investigated and found to give a satisfactory fit, with a reasonable ordered moment. Other cubic magnetic space groups consistent with the propagation vectors can be ruled out by systematic absence considerations. 

\begin{figure}[!b]
\begin{center}
\includegraphics[width=0.55\textwidth]{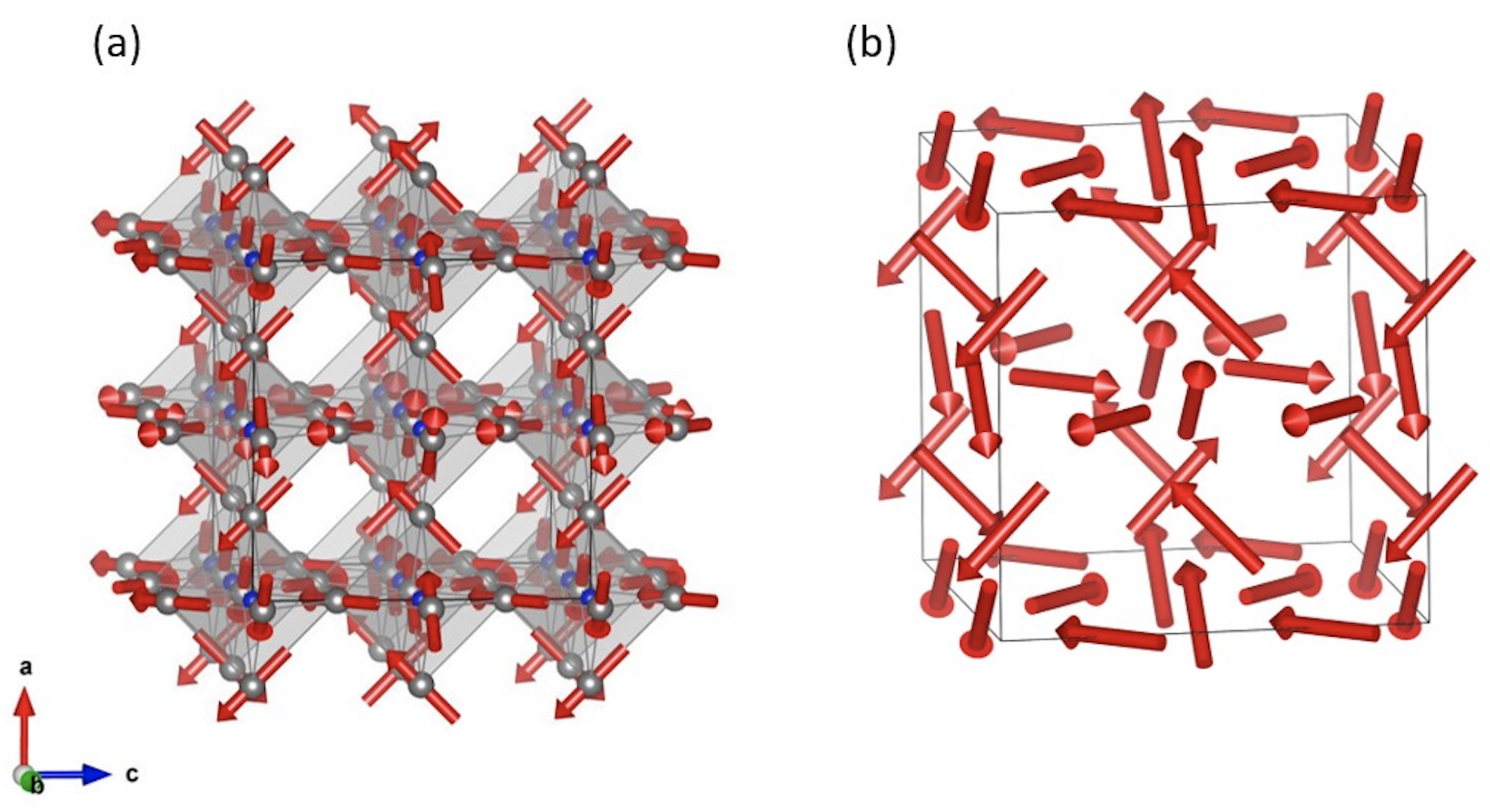}
\end{center}
\caption{The zero-field magnetic structure of Eu$_{3}$PbO (upper panel) and spin-only representation (lower panel).}
  \label{fig:fig3a}
\end{figure}

\begin{figure}[!b]
\begin{center}
\includegraphics[width=0.55\textwidth]{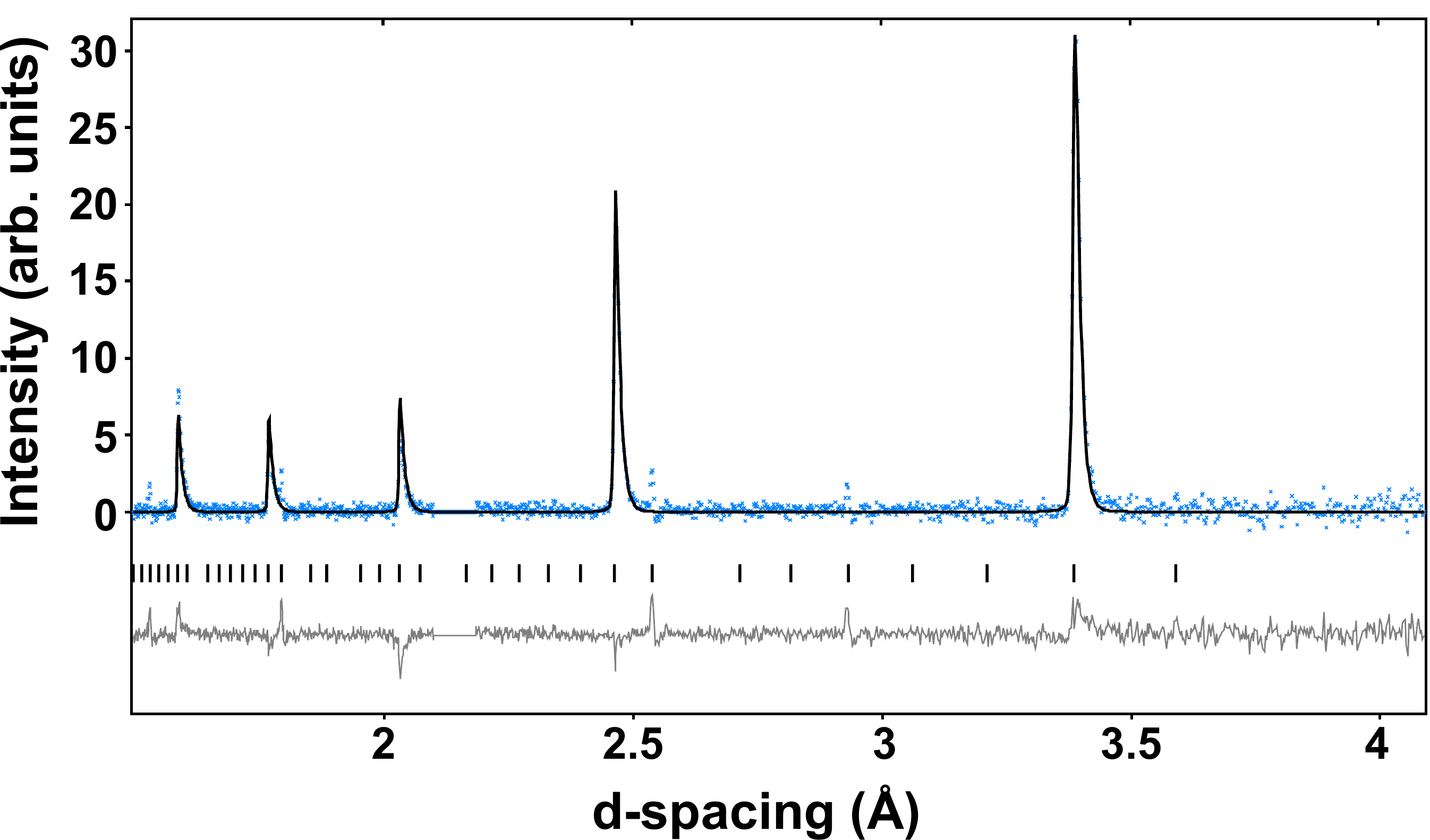}
\end{center}
\caption{A portion of the Rietveld refinement for the $P_{I}a\bar{3}$ structure (90$^{\circ}$ banks 3 \& 8 WISH, magnetic only subtracted pattern).}
  \label{fig:fig3}
\end{figure}

This structure can be rationalized with two Eu-Eu interactions: a strong FM interaction through the oxygen and an AFM direct exchange along the octahedral edges that leads to a 120$^{\circ}$ structure, as shown in Fig.~\ref{fig:fig3a}.  The refinement results are tabulated in Table~\ref{tab:tabc}, and fitted profiles are shown in Fig.~\ref{fig:fig3}. Since this model gives a reasonable ordered moment for a single Eu site, and with the current data we cannot distinguish between single- and multi-$k$ order, we have taken the multi-$k$ $P_{I}a\bar{3}$ model as the most appropriate for the zero-field state, since it can be easily rationalized using the known interactions of Eu$^{2+}$. 

\begin{table}[t!]
  \centering
   \caption{The results of the Rietveld refinement of the $1.5\,$K 0 T Eu$_3$PbO data in magnetic space group $P_{I}a\bar{3}$, in the standard setting where a=$2a_p$, b=$2a_p$, c=$2a_p$, and $a_p$ is the parent lattice constant $a_p =10.15638(10)$\AA. }
\vspace{0.3cm}
\begin{ruledtabular}
  \begin{tabular}{ccccccc}
   Site & x & y & z & $m_{x}$ & $m_{y}$ & $m_{z}$  \\ 
    &  &  &  & ($\mu_{B}$/Eu$^{2+}$)& ($\mu_{B}$/Eu$^{2+}$) & ($\mu_{B}$/Eu$^{2+}$) \\ 
\hline
Eu1 & 0 & 0.5 & 0.25 & 0 & -4.769 $\pm$0.011 & 4.769 $\pm$0.011  \\  
  \end{tabular}
  \end{ruledtabular} 
  \label{tab:tabc}
\end{table}

\subsection{Applied field magnetic structures} 

\begin{figure}[t]
\begin{center}
\includegraphics[width=0.55\textwidth]{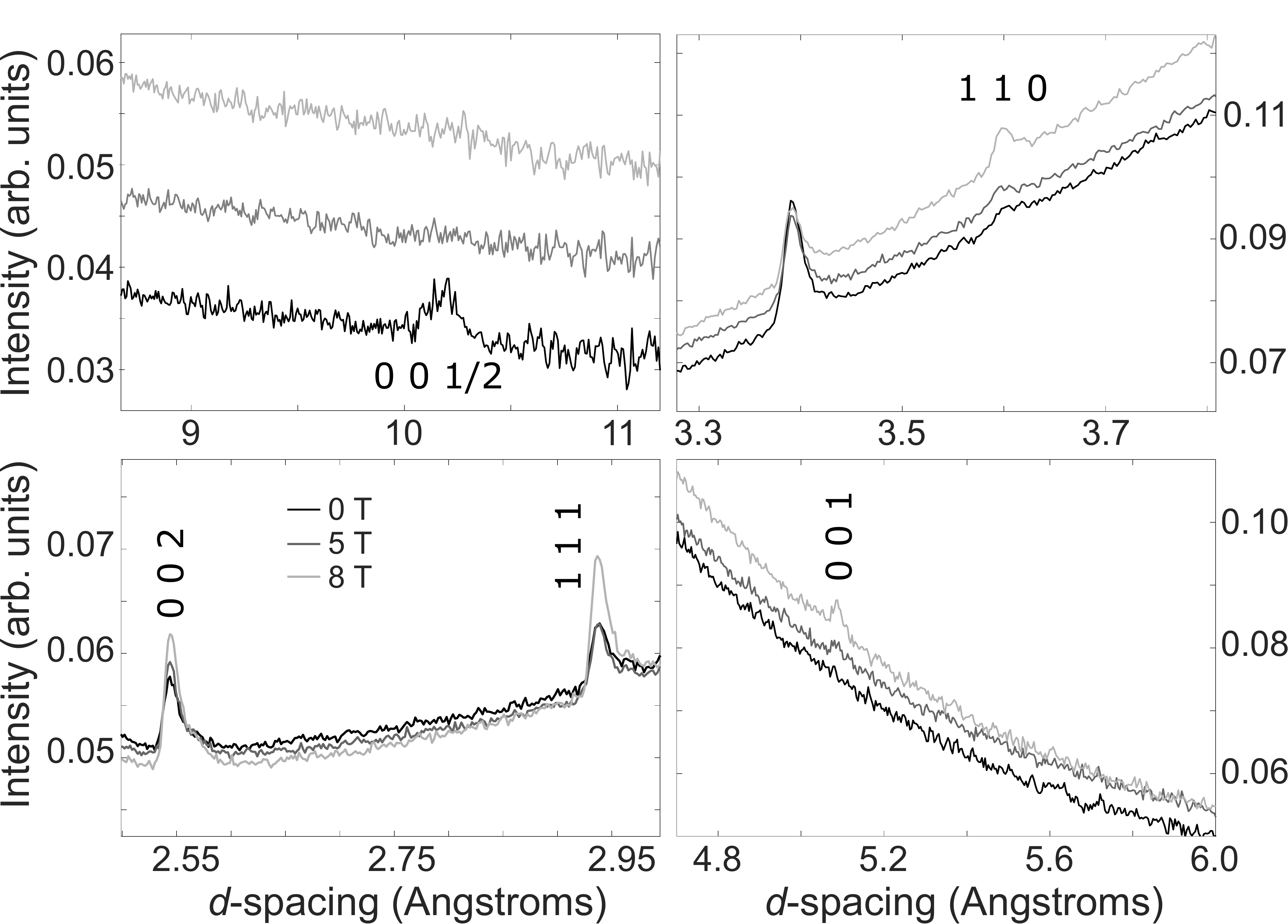}
\end{center}
\caption{Peak intensity changes as a function of applied field at $T=1.5\,$K. Reflection labels are in the notation of the parent unit cell.}
  \label{fig:fig4}
\end{figure}

\begin{figure}[!t]
\begin{center}
\includegraphics[width=0.55\textwidth]{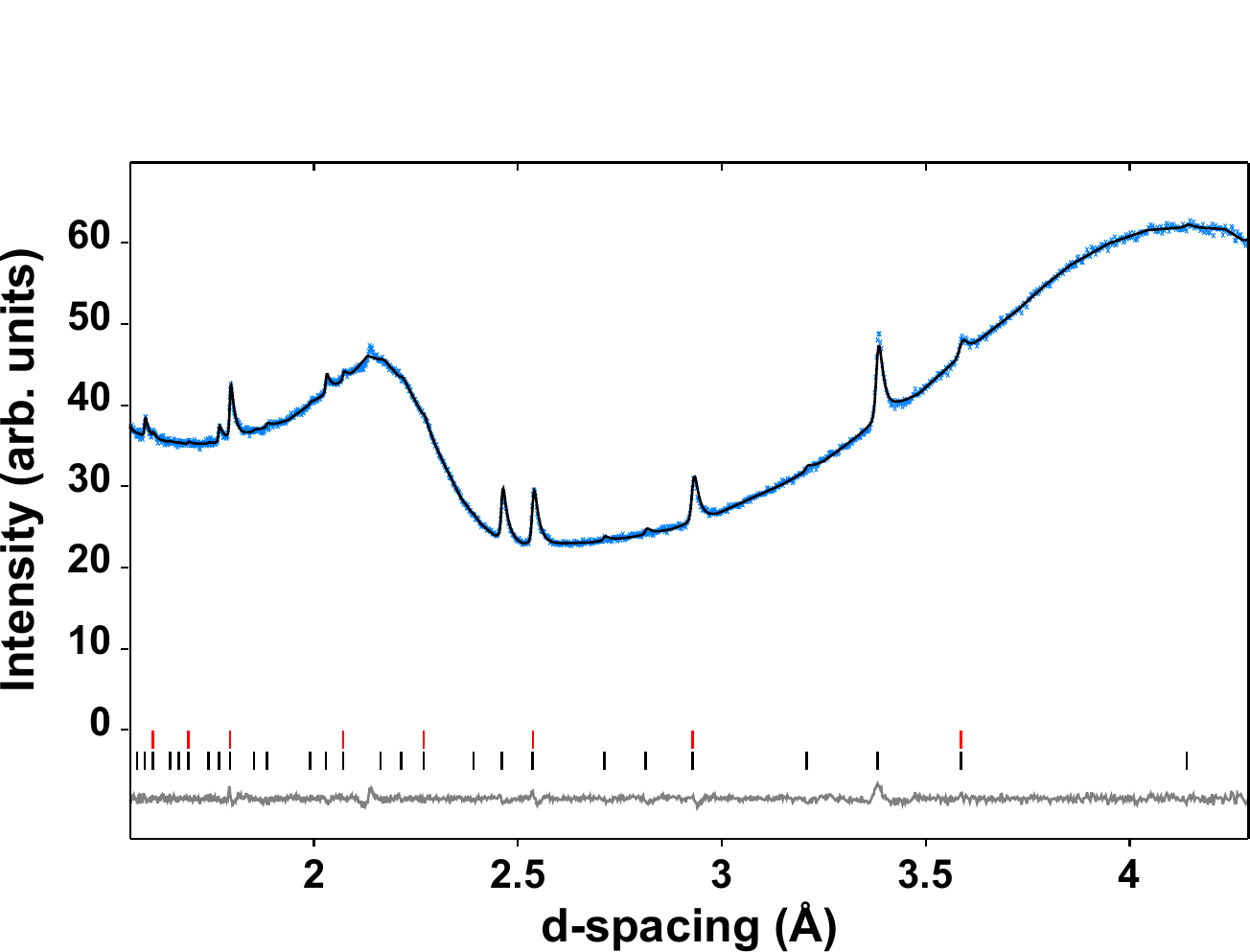}
\end{center}
\caption{A portion of the Rietveld refinement for the 1.5 K, 5 T structure in $P4/mm^{\prime}m^{\prime}$ (90$^{\circ}$ bank 3 WISH). The unindexed peak at 2.14 \AA{} is from vanadium, the substantial background is from the quartz ampoule used to contain the sample.}
  \label{fig:fig6}
\end{figure}

\begin{figure}[!t]
\begin{center}
\includegraphics[width=0.55\textwidth]{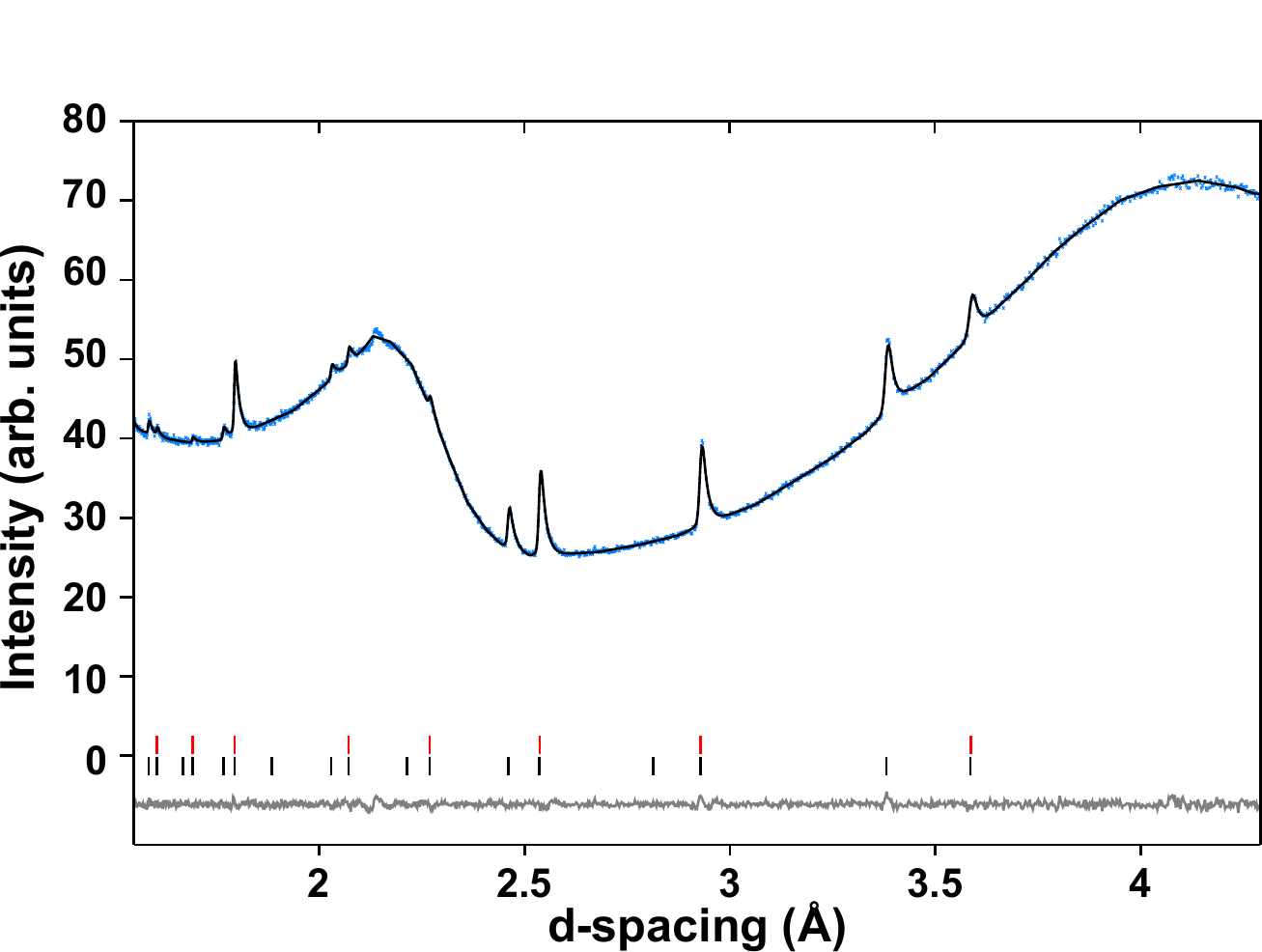}
\end{center}
\caption{A portion of the Rietveld refinement for the 1.5 K, 8 T structure in $Pm^{\prime}m^{\prime}m$ (90$^{\circ}$ bank 3 WISH). The unindexed peak at 2.14 \AA{} is from vanadium, the substantial background is from the quartz ampoule used to contain the sample.} 
 \label{fig:fig7}
\end{figure}

The $P_{I}a\bar{3}$ model allows for a simple evolution of the structure with applied magnetic field, with the magnetic structure losing one arm of the star of $\bf{k}$ at each magnetic transition (we consider field applied along the $c$-axis). The addition of propagation vector (0,0,0) captures the spontaneous ferromagnetic moment. For the first transition at 5 T the propagation vector (0,0,$\frac{1}{2}$) is lost and a model in $P$4$/mm^{\prime{}}m^{\prime{}}$ with five Eu sites, constrained to have equal moment, gives an excellent fit and reasonable ordered moment (see results in Table \ref{tab:tab3} and Fig.~\ref{fig:fig6}). At the second transition a further arm of the star, (0,$\frac{1}{2}$,0) is lost, resulting in $Pm^{\prime{}}m^{\prime{}}m$ with 5 sites. With these sites again constrained to have equal moments an excellent fit is obtained, the results of which are summarised in Table~\ref{tab:tab4} and Fig.~\ref{fig:fig7}. The ordered moment obtained is slightly smaller than the full ordered moment for Eu$^{2+}$ but considering the magnetic anisotropy present, it is possible that there is a certain amount of phase coexistence in this higher-field region, which the current powder measurements are relatively insensitive to. Therefore the overall phase transition sequence can be suggested to be $P_{I}a\bar{3}\rightarrow{}P$4$/mm^{\prime{}}m^{\prime{}}\rightarrow{}Pm^{\prime{}}m^{\prime{}}m$ assuming an applied field along the (0,0,1) direction. This is justified by the impossibility of determining moment direction in a metrically cubic material from powder data.
The magnetic structures of the Ferri-II and Ferri-III phases, decomposed into ferromagnetic and antiferromagnetic components, are shown in Fig.~\ref{new_Fig_S7}.

Overall, we find this sequence based on the multi-$k$ model to be the most phenomenologically reasonable and consistent with the known interactions within the system. It is also consistent with all physical property data, including the magnitudes of the jumps in magnetization at each magnetic transition. It should be noted that single crystal measurements will be required to fully confirm the magnetic space groups and exact moment directions in all magnetic phases, but in the meantime a reasonable working solution has been obtained.

\begin{table}[h]
  \centering
   \caption{The results of the Rietveld refinement of the $1.5\,$K, 5 T Eu$_3$PbO data in magnetic space group $P4/mm^{\prime}m^{\prime}$, a=$2a_p$, b=$2a_p$, c=$a_p$, where $a_p$ is the parent lattice constant. }
\vspace{0.3cm}   
\begin{ruledtabular}
  \begin{tabular}{ccccccc}
   \hline 
   Site & x & y & z & $m_{x}$ & $m_{y}$ & $m_{z}$  \\ 
    &  &  &  & ($\mu_{B}$/Eu$^{2+}$)& ($\mu_{B}$/Eu$^{2+}$) & ($\mu_{B}$/Eu$^{2+}$)\\
\hline
Eu1 & 0 & 0.5 & 0 & 0 & 0  & 6.45 $\pm$0.06  \\  
Eu2 & 0.25 & 0 & 0.5 & 0 &  0  & -6.45 $\pm$0.06 \\
Eu3 & 0.75 & 0.5 & 0.5 & 0 &  0  & 6.45 $\pm$0.06 \\ 
Eu4 & 0.5 & 0.5 & 0.0 & 0 &  0  & 6.45 $\pm$0.06 \\
Eu5 & 0 & 0 & 0 & 0 &  0  & -6.45 $\pm$0.06 \\
  \end{tabular} 
  \end{ruledtabular}
  \label{tab:tab3}
\end{table}

\begin{table}[h]
  \centering
   \caption{The results of the Rietveld refinement of the $1.5\,$K, 8 T Eu$_3$PbO data in magnetic space group $Pm^{\prime}m^{\prime}m$, a=$2a_p$, b=$a_p$, c=$a_p$, where $a_p$ is the parent lattice constant. }
\vspace{0.3cm}
\begin{ruledtabular}
  \begin{tabular}{ccccccc}
   Site & x & y & z & $m_{x}$ & $m_{y}$ & $m_{z}$  \\ 
    &  &  &  & ($\mu_{B}$/Eu$^{2+}$)& ($\mu_{B}$/Eu$^{2+}$) & ($\mu_{B}$/Eu$^{2+}$)\\
\hline
Eu1 & 0 & 0.5 & 0.5 & 0 & 0 & 5.08 $\pm$0.06  \\ 
Eu2 & 0.25 & 0 & 0.5 & 0 & 0 & -5.08 $\pm$0.06  \\ 
Eu3 & 0.25 & 0.5 & 0 & 0 &0 & -5.08 $\pm$0.06  \\ 
Eu4 & 0.75 & 0 & 0.5 & 0 &0 & 5.08 $\pm$0.06 \\ 
Eu5 & 0.75 & 0.5 & 0 & 0 & 0 & 5.08 $\pm$0.06  \\ 
  \end{tabular}
  \end{ruledtabular} 
  \label{tab:tab4}
\end{table}

\clearpage{}

\refstepcounter{section}
\section*{Supplementary Note III: Tight-binding model for $\textrm{Eu$\textsubscript{3}$PbO}$} \label{simple model} \label{TB_model} \label{supp_note_one}

To construct a   tight-binding model for Eu$_3$PbO we follow along the lines of the works by Kariyado and Ogata~\cite{kariyadoJPSJ12}
and Chiu \textit{et al.}~\cite{chiu_PRB_17}.
In Ref.~\cite{chiu_PRB_17} a nine-band model for Ca$_3$PbO with three Pb-$p$ orbitals and six 
Ca-$d$ orbitals was constructed.
This model captures the low-energy physics of Ca$_3$PbO faithfully. In particular, it exhibits
six gapped Dirac cones along 
 the $\Gamma-X$ direction with a non-zero mirror Chern number,
in full agreement with the \textit{ab-initio} DFT calculations. 
In the following we describe how this model can be
adapted to the case of Eu$_3$PbO, both for the paramagnetic
phase and the magnetically ordered phases.

\vspace{1.0 cm} 
 
\subsection{Paramagnetic phase}

The paramagnetic phase of Eu$_3$PbO 
can be described by the same model as in Ref.~\cite{chiu_PRB_17}, albeit with different parameter values,
since its band structure is qualitatively similar to Ca$_3$PbO. 
In the absence of spin-orbit coupling, this tight-binding model is written as
$\mathcal{H}_{\textrm{PM}} = \sum_{\bf k}   \psi^{\dag}_{\bf k}  H_{\textrm{PM}}  ( {\bf k} ) \psi_{\bf k} $
with the nine-component spinor
\begin{align*}
\psi_{\bf k}=
(&\mathrm{Pb}_{p_x},\quad\quad\mathrm{Pb}_{p_y}, \quad\quad\mathrm{Pb}_{p_z}, \\
&\mathrm{Eu}^1_{d_{y^2-z^2}},\quad \mathrm{Eu}^2_{d_{z^2-x^2}}, \quad\mathrm{Eu}^3_{d_{x^2-y^2}}, \\
&\mathrm{Eu}^1_{d_{yz}},\quad\quad\mathrm{Eu}^2_{d_{zx}},\quad\quad\mathrm{Eu}^3_{d_{xy} } )^{\textrm{T}}
\end{align*}
and the $9 \times 9$ matrix $H_{\textrm{PM}} ( {\bf k} )$ with block form
\begin{eqnarray} \label{ham_wo_soc}
H_{\textrm{PM}} ( {\bf k} )
=
\begin{pmatrix}
H_p & V_{dp}^u & V_{dp}^l \cr
{V_{dp}^u}^{\dag} & H_d^u & 0 \cr
{V_{dp}^l}^\dag & 0 & H_d^l \cr
\end{pmatrix} .
\end{eqnarray}
The blocks of $H_{\textrm{PM}}   ( {\bf k} )$ are given by
\begin{eqnarray}
H_p
=
\begin{pmatrix}
e_p - 2 t_{pp} c_{2x}  & 0 & 0   \cr
0 & e_p - 2 t_{pp} c_{2y}  & 0  \cr
0 & 0 & e_p - 2 t_{pp} c_{2z}  \cr
\end{pmatrix} ,
\end{eqnarray}
\begin{eqnarray}
H_d^u
=
\begin{pmatrix}
e_d & - 4 t_{dd} c_x c_y & - 4 t_{dd} c_z c_x \cr
- 4 t_{dd} c_x c_y &  e_d & - 4 t_{dd} c_y c_z \cr
- 4 t_{dd} c_z c_x & - 4 t_{dd} c_y c_z & e_d  \cr
\end{pmatrix},   
\end{eqnarray}
and $H_d^l=e_d\bI_3$,  
with $\bI_3$ the 3$\times$3 identity matrix.
The coupling matrices between Pb-$p$ and Eu-$d$ orbitals read
\begin{eqnarray}
V_{dp}^u
&=&
4 i t_{pd}
\begin{pmatrix}
0 & c_z s_x & - c_y s_x  \cr
- c_z s_y & 0 & c_x s_y \cr
c_y s_z & - c_x s_z& 0 \cr 
\end{pmatrix}, 
\nonumber\\
V_{dp}^l
&=&
4 i t_{pd}
\begin{pmatrix}
 0 & c_x s_z & c_x s_y \cr
 c_y s_z & 0 & c_y s_x \cr
 c_z s_y & c_z s_x & 0 \cr 
\end{pmatrix},
\end{eqnarray}
where we have used the abbreviations
$
c_i = \cos \frac{ k_i }{2}
$,
$
s_i = \sin \frac{ k_i }{2}
$,
and
$
c_{2i} 
= \cos k_i $.

To lowest order, spin-orbit coupling enters as an on-site term in the tight-binding Hamiltonian~\eqref{ham_wo_soc}. 
For the Pb-$p$ orbitals the on-site spin-orbit coupling reads
 $\sum_{\bf k} \psi^{\dag}_p ({\bf k} )  H^p_{\mathrm{SO}} ( {\bf k} )  \psi^{\ }_p ({\bf k} ) $ with
the spinor 
\begin{eqnarray}
\psi^{\ }_p ({\bf k} ) = ( \mathrm{Pb}^{\uparrow}_{p_x} ,  \mathrm{Pb}^{\uparrow}_{p_y},  \mathrm{Pb}^{\uparrow}_{p_z}  ,  \mathrm{Pb}^{\downarrow}_{p_x} ,  \mathrm{Pb}^{\downarrow}_{p_y},  \mathrm{Pb}^{\downarrow}_{p_z}   ) \nonumber
\end{eqnarray}
and
\begin{eqnarray}
H^p_{\mathrm{SO}} ( {\bf k} ) 
=
\frac{\lambda_p}{2} 
\begin{pmatrix}
0 & - i & 0 & 0 & 0 & 1 \cr
i & 0 & 0 & 0 &0 & - i \cr
0 & 0 & 0 & - 1 & \; i \; & 0 \cr
0 & 0 & -1 & 0 & \; i \;  & 0 \cr
0 & 0 & - i & - i & 0 & 0 \cr
1 & i & 0 & 0 &0 & 0 \cr
\end{pmatrix} . \nonumber
\end{eqnarray}
The on-site spin-orbit coupling for the Eu-$d$ orbitals is $\sum_{\bf k} \psi^{\dag}_d ({\bf k} )  H^d_{\mathrm{SO}} ( {\bf k} )  \psi^{\ }_d ({\bf k} ) $ with
the vector 
\begin{eqnarray} 
&& \psi_d ( {\bf k} ) =
\nonumber\\
&& (  \mathrm{Eu}^{1, \uparrow} _{d_{y^2-z^2}} ,  
 \mathrm{Eu}^{2, \uparrow}_{d_{z^2-x^2}},
 \mathrm{Eu}^{3, \uparrow}_{d_{x^2-y^2}} ,
  \mathrm{Eu}^{1, \downarrow} _{d_{y^2-z^2}} ,  
 \mathrm{Eu}^{2, \downarrow}_{d_{z^2-x^2}},
 \mathrm{Eu}^{3, \downarrow}_{d_{x^2-y^2}} ,
\nonumber\\
&&
  \mathrm{Eu}^{1, \uparrow}_{d_{yz}},
  \mathrm{Eu}^{2, \uparrow}_{d_{zx}},
 \mathrm{Eu}^{3, \uparrow}_{d_{xy} } ,
  \mathrm{Eu}^{1, \downarrow}_{d_{yz}},
  \mathrm{Eu}^{2, \downarrow}_{d_{zx}},
 \mathrm{Eu}^{3, \downarrow}_{d_{xy} } ,
 )^{T}, 
 \nonumber
\end{eqnarray}
and
\begin{align}
 H^{d}_{\mathrm{SO}} ( {\bf k} ) 
 =&\lambda_d  \tau_y \otimes \Big \{
 \sigma_x \otimes 
\begin{pmatrix}
  1 & 0 & 0\cr
  0 & 0 & 0 \cr
  0 & 0 & 0 \cr
 \end{pmatrix} \nonumber \\
&+  \sigma_y \otimes 
\begin{pmatrix}
  0 & 0 & 0\cr
  0 & 1 & 0 \cr
  0 & 0 & 0 \cr
   \end{pmatrix} 
  + \sigma_z \otimes 
\begin{pmatrix}
  0 & 0 & 0\cr
  0 & 0 & 0 \cr
  0 & 0 & 1 \cr
 \end{pmatrix}
\Big \},
 \end{align}
where $\tau_\beta$  and $\sigma_\alpha$ operate in the $d$-orbital ($d_{x_i^2-x^2_j}$ and $d_{x_ix_j}$) and spin ($\uparrow$ and $\downarrow$) 
degree of freedom, respectively. 
Combining these spin-orbit coupling terms with Eq.~\eqref{ham_wo_soc}, we obtain the full tight-binding Hamiltonian for 
the paramagnetic phase of Eu$_3$PbO
\begin{eqnarray} \label{ham_with_SOC}
H_{\textrm{PM}}^{\textrm{tot}}  ( {\bf k} ) 
=
\begin{pmatrix}
H_p^{\textrm{tot}}( {\bf k} ) & V_{\textrm{tot}} ( {\bf k} ) \cr
V^{\dag}_{\textrm{tot}} ( {\bf k}\hat{E})  &  H_d^{\textrm{tot}}  ( {\bf k} ) \cr
\end{pmatrix} + 
\mu \bI_{18}, 
\end{eqnarray}
with 
\begin{align} \label{def_HpTot}
H_p^{\textrm{tot}} ( {\bf k} ) 
=&
\begin{pmatrix}
H_p & 0 \cr
0 & H_p \cr
\end{pmatrix}
+
 H^p_{\mathrm{SO}} ( {\bf k} ) ,   \\
 H_d^{\textrm{tot}}  ( {\bf k} ) 
=&
\begin{pmatrix}
\sigma_0\otimes H_d^u & 0 \cr
0 & \sigma_0 \otimes H_d^l \cr
\end{pmatrix}
 +
 H^{d}_{\mathrm{SO}} ( {\bf k} ),  \nonumber
\end{align}
and 
\begin{eqnarray}
V_{\textrm{tot}} ( {\bf k} )
=
\begin{pmatrix}
\sigma_0\otimes V^u_{dp} &   \sigma_0\otimes V^l_{dp}   \cr
\end{pmatrix}. \nonumber
\end{eqnarray} 
The outermost grading of $H_p^{\textrm{tot}}$ and $\sigma_0$ in 
Eq.~\eqref{def_HpTot} corresponds to the spin grading.
In Eq.~\eqref{ham_with_SOC} a diagonal term $\mu \bI_{18}$ for the chemical potential has been added.

We have determined the nine parameters of the above tight-binding model
by a fit to the \textit{ab-initio} DFT band structure, which yields
\begin{eqnarray}
&&
e_p = 0.0, 
\quad
e_d = 2.4,
\quad
t_{pp} = -0.4,
\quad
t_{dd} = -0.4,
\nonumber\\
&&
t_{pd} = -0.4,
\quad
\lambda_p =0.75,
\quad
\lambda_d = 0.07,
\quad
\mu = 0.87 .
\nonumber
\end{eqnarray}

\subsection{Magnetic phases}

To describe the Eu magnetic moments we introduce the magnetic splitting derived by the DFT calculations
as Zeeman terms $H_{\text{Zee}, n}$ for each orbital $n$  into the tight-binding model as
\begin{eqnarray}
 H_{\text{Zee}, n} = ( 
 \begin{pmatrix}
\sigma_x \cr
\sigma_y \cr
\sigma_z \cr
\end{pmatrix}
 \cdot {\bf B}_n)  \ket{n}\bra{n}  ,
\end{eqnarray}
where $\ket{n}\bra{n} $ is the projector onto the orbitals $n$ without spin degree of freedom.
The Pauli matrices $\sigma_i$ describe the spin degree of freedom and ${\bf B}_n$ is the magnetic splitting energy. 

The magnitude of the splitting energy ${\bf B}_n$ for the different orbitals is determined from the DFT calculations of the ferromagnetic phase. 
For the other magnetic phases we then assume that the splitting vector ${\bf B}_n$
reorients according to the respective magnetic structure, but does not change its magnitude.
We have checked that this procedure leads to a tight-binding band structure that is qualitatively similar to the DFT electronic bands. 
 
\subsubsection{Ferromagnetic phase}

In the ferromagnetic phase all moments are aligned collinearly. Therefore,  ${\bf B}_n$ points in the same direction ${\bf \hat{B}}$  at all sites
and we can write ${\bf B}_n = B_n {\bf \hat{B}}$. By comparing to DFT calculations, we find that the magnetic splitting $B_n$
of the different orbitals are 
\begin{eqnarray}
 B_{\text{Eu}}  = 0.43 \text{ eV}, 
\quad
 B_{\text{Pb}}  = -0.035 \text{ eV}, 
\end{eqnarray}
which corresponds to half   the total  energy splitting, as read out from the DFT band structures. 

In the ferromagnetic phase the magnetization direction ${\bf \hat{B}}$ can be easily aligned by the external field. 
As explained in the main text, the topology of the electronic bands changes with magnetization direction. In particular, 
the position of the Weyl points, both in energy and momentum, as well as their multiplicities depend strongly 
on the magnetization direction. This is shown in Table~\ref{table_WeylpointsInFM} for the three magnetization directions
[100], [110], and [111], see also Fig.~\ref{mFig2} in the main text. 
Note that the Weyl points W1 only exist   for the [110] and [111] magnetizations, but are absent for the [100] magnetization. 
For the [110] magnetization W1 has multiplicity four, i.e., there are two pairs of Weyl points, one close to the $k_x$ axis and another
one close to the $k_y$ axis. For the [111] magnetization there are six Weyl points W1, as the symmetry is higher. I.e., there is
one pair of Weyl points close to each of the three mains axis $k_x$, $k_y$, and $k_z$. 
Similarly, the multiplicity of the Weyl points W2 and W3 is only four for the [110] magnetization, while it is six for the [111] magnetization. 
For the [110] magnetization the Weyl points W4 have also multiplicity four, but now they are located close to the $k_z$ axis, one pair with positive $k_z$
and one pair with negative $k_z$.

The ferromagnetic phases with [100] and [110] magnetization exhibit also nodal lines. These nodal lines are located in the plane perpendicular 
to the magnetization direction, i.e. in the $k_y k_z$-plane and in the [110]-plane, respectively.
These nodal lines are protected by mirror symmetry and by a quantized $\pi$-Berry phase.

\begin{table}[t!]
\centering
\begin{ruledtabular}
\begin{tabular}{ccclcc}
phase & position    &  E (eV)  & type & top.~inv. & $\#$ \\ 
\hline
FM [100]& {\footnotesize $(0.18, 0, 0)$ }& -0.16 & WP (W2) & Chern & 2 \\
FM [100] &{\footnotesize $(0.12, 0, 0)$ } & -0.33 & WP (W3) & Chern & 2 \\
FM [100] & {\footnotesize $k_y k_z$-plane}  & -0.02 &  Line (L1) & Berry & 1 \\
FM [100] & {\footnotesize $k_y k_z$-plane} & -0.31 &  Line (L2) & Berry & 1 \\
\hline
FM [110] & {\footnotesize $(0.23, 0.015, 0)$} & 0.06 & WP (W1) & Chern & 4 \\
FM [110]& {\footnotesize $(0.19, -0.001, 0)$ }& -0.12 & WP (W2) & Chern & 4 \\
FM [110] &{\footnotesize $(0.12, 0, 0) $ } & -0.32 & WP (W3) & Chern & 4 \\
FM [110] & {\footnotesize $( 0.003,  0.003 ,0.13)$} & -0.31 & WP (W4) & Chern & 4 \\
FM [110] & {\footnotesize [110]-plane} & -0.02 & Line (L1) & Berry  & 1 \\
\hline
FM [111] & {\footnotesize $(0.23, 0.008, 0.008)$} & 0.05 & WP (W1) & Chern & 6 \\
FM [111] & {\footnotesize $(0.2,-0.001,-0.001)$ }& -0.09 & WP (W2) & Chern & 6 \\
FM [111] &{\footnotesize $(0.13, 0, 0)$ } & -0.31 & WP (W3) & Chern & 6 
\end{tabular}
\end{ruledtabular}
\caption{
\textbf{Weyl points of the ferromagnetic phase for different field orientations.}
This table lists the positions and energies of the topological band crossings in the first Brillouin zone (BZ)
for the ferromagnetic phase with magnetization in [100], [110], and [111]   direction (FM [100], FM [110], and FM [111], respectively).
 The positions of the band crossings ${\bf k} = ( k_x, k_y, k_z)$ are given in 
units of $2 \pi/ a_{i}$, where $a_i$ denotes the lattice constant of the respective real space direction. All energies are given in eV relative to the Fermi energy. 
The type of band crossing is indicated in the fourth column, while the fifth column 
states the topological invariant that protects the crossings.
The last column gives the multiplicity of the crossings, i.e., the number of symmetry related crossings
at the same energy.
\label{table_WeylpointsInFM} 
}
\end{table} 

\subsubsection{Antiferromagnetic phase}

The magnetic space group of the AFM-I phase is $P_Ia\bar{3}$ (No.~205.36, type IV).
The unit cell is eightfold enlarged as compared to the paramagnetic phase. I.e., it is doubled
in each of the three main axes $x$, $y$, and $z$. 
This leads to an eightfold back folding of the bands, and hence the tight-binding model
of the AFM-I phase has 8 $\times$ 18 = 144 bands (including spin).
Correspondingly, there are eight times more orbitals in the tight-binding model,
leading to an 144 $\times$ 144 tight-binding Hamiltonian.
The hopping parameters for this enlarged Hamiltonian can be determined in an automatized fashion
directly in momentum space from the tight-binding model of the paramagnetic phase, Eq.~\eqref{ham_with_SOC}.
For that purpose, we first perform a unitary transformation of Hamiltonian~\eqref{ham_with_SOC} in order to simplify its momentum dependence.
This transformation amounts to multiplying 
the Pb-$p$ orbitals by $\mathrm{e}^{i (k_x + k_y + k_z)/2}$, 
the $\mathrm{Eu}^1$-$d$ orbitals by $\mathrm{e}^{i k_x /2}$, 
the $\mathrm{Eu}^2$-$d$ orbitals by $\mathrm{e}^{i k_y /2}$, and 
the $\mathrm{Eu}^3$-$d$ orbitals by $\mathrm{e}^{i k_z /2}$. 
With this, all the terms in Eq.~\eqref{ham_with_SOC}
of the form  $\mathrm{e}^{i k_n /2}$ are transformed into terms with  $\mathrm{e}^{i k_n }$ 
or terms that are  independent of $k_n$. 
Now, we can start to construct the tight-binding Hamiltonian for the AFM-I phase,
which has a block structure with 8 $\times$ 8 blocks, where each block is an 18 $\times$ 18 matrix, i.e., 
\begin{align} \label{Ham_AFM}
H_{\mathrm{AFM}}({\bf k}) = 
\begin{pmatrix}
  H_{000}& H_{x}({\bf k}) &  H_{y}({\bf k}) &  H_{xy}({\bf k}) &  H_{z}({\bf k}) &  H_{xz}({\bf k}) &  H_{yz}({\bf k}) & 0  \cr
  & H_{100} &  H_{y}^x({\bf k}) & H_{y}({\bf k}) & H_x^z({\bf k}) &  H_{z}({\bf k})  & 0 & H_{yz}({\bf k})  \cr 
  &  & H_{010} & H_{x}({\bf k}) & H_z^y({\bf k}) & 0 &  H_{z}({\bf k}) &  H_{xz}({\bf k})  \cr
  &  &  & H_{110} & 0 & H_z^y({\bf k}) &  H_x^z({\bf k}) &  H_{z}({\bf k})  \cr
  &  &  &  & H_{001} & H_{x}({\bf k}) & H_{y}({\bf k}) &  H_{xy}({\bf k}) \cr
  &  & H.C.  &  &  & H_{101} & H_{y}^x({\bf k}) & H_{y}({\bf k})  \cr
  &  &  &  &  &  & H_{011} &  H_{x}({\bf k}) \cr
  &  &  &  &  &  &  & H_{111}  \cr
\end{pmatrix}.
\end{align}
The 18 $\times$ 18 matrices $H_{abc}$ (with $a,b,c \in \{0,1\}$) on the diagonal describe
hoppings within each of the eight paramagnetic unit cells.
The off-diagonal entries $H_{x}$, $H_{y}$, $H_{z}$, $H_{xy}$, $H_{xz}$, $H_{yz}$, $H_y^x$, $H_x^z$, and $H_z^y$  describe
hoppings that connect different paramagnetic unit cells. These hopping terms are modified by exponential factors $\mathrm{e}^{i k_n}$,
since they connect different paramagnetic unit cells.

So far, Eq.~\eqref{Ham_AFM} represents just an artificial increase of the Hamiltonian, that trivially leads to folded bands. 
But now, we introduce the magnetic splitting ${\bf B}_i =  B_{\text{Eu}}  {\bf \hat{B}}_i$ due to the antiferromagnetically ordered Eu moments. These
splitting energies are added to the diagonal blocks $H_{abc}$ in Eq.~\eqref{Ham_AFM}
and have all the same magnitude 
\begin{eqnarray}
B_{  \text{Eu}}  = 0.45 \text{ eV}.
\end{eqnarray}
The orientation of the ${\bf \hat{B}}_n$ vectors on the different Eu sites is determined 
by the AFM ordering pattern, as given in the main text. 
We can implement this pattern in the following way
\begin{align}
{\bf \hat{B}}_{\text{Eu}^1} &= \frac{1}{\sqrt{2} } (0,(-1)^a,(-1)^b ),  
 \nonumber \\
{\bf \hat{B}}_{\text{Eu}^2} &= \frac{1}{\sqrt{2} } ((-1)^c,(-1)^a,0 ),
 \nonumber \\
 {\bf \hat{B}}_{\text{Eu}^3} &= \frac{1}{\sqrt{2} } ((-1)^c, 0 ,(-1)^b ),
\end{align}
where the indices $a,b,c \in \{0,1\}$ label the eight different paramagnetic unit cells.

We note that due to backfolding, the bands in the AFM-I phase exhibit band crossings at the time-reversal invariant
momenta $X$, $Y$, and $Z$ (D2 in Fig.~\ref{mFig2}(g) of the main text).
Hybridization at these points is strongly suppressed by symmetry.
To explain this, we first note that the bands at the $X$, $Y$, and $Z$ points have almost exclusively Pb-$p$ orbital character,
with only very small admixtures of $Eu$-d orbital character. Now, due to the combination of inversion and non-symmorphic time-reversal symmetry $\widetilde{\mathcal{T}}$,
the Pb-$p$ orbitals cannot carry a finite magnetic moment. Hence, the splitting of the bands at  the $X$, $Y$, and $Z$ points, is negligibly small,  leading
to nearly  gapless Dirac points. 
From our DFT and tight-binding calculations we find that the gap of these Dirac points
is indeed small, namely smaller than $~1$ meV.
This small  gap is caused by a very small, but finite, admixture of 
Eu-$d$ orbital character.

 \clearpage
 
  \newpage

\refstepcounter{section}
\section*{Supplementary Note IV: \\ Topological Invariants, Anomalous Hall Conductivity, and Surface States} \label{supp_note_two}

Here, we explain how the Chern numbers, the anomalous Hall conductivities, and the surface states are computed.  
We also give a detailed symmetry analysis of the anomalous Hall conductivity tensor and determine the   
contributions of the different Weyl points to the anomalous Hall conductivity. 

\vspace{1 cm}

\subsection{Chern number}

The numerical computation of the Chern number follows the approach of Fukui \textit{et al.} \cite{2005_Fukui}. 
This approach uses the U(1) link variable $U_{\mu}({\bf k}_l)$ on a discretized Brillouin zone to define the lattice 
field strength $\tilde{F}_{12}({\bf k}_l)$, which represents a small Wilson loop for one eigenstate of the Hamiltonian. 
The Chern number is then the sum over all occupied bands and all points in the two-dimensional Brillouin zone. 
The link variable and the lattice field strength are defined as
\begin{eqnarray}
 U_{\mu}({\bf k}_l) 
 &=&
  \frac{\bra{n({\bf k}_l)}n({\bf k}_l+ \hat{\mu}) \rangle}{\left| \bra{n({\bf k}_l)}n({\bf k}_l+ \hat{\mu}) \rangle \right|}
\end{eqnarray}
and
\begin{eqnarray}
 \tilde{F} _{12}({\bf k}_l) 
 &=&
  \ln\left(  U_{1}({\bf k}_l)  U_{2}({\bf k}_l + \hat{1})  U_{1}({\bf k}_l + \hat{2})^{-1} U_{2}({\bf k}_l)^{-1} \right),
  \nonumber
\end{eqnarray}
respectively, where $ \ket{{ n (\bf k}_l )}$ is the eigenstate for the nth non-degenerate band at point ${\bf k}_l$ in the Brillouin zone. 
$\hat{\mu}$ as well as the explicit versions $\hat{1}$ and $\hat{2}$ denote steps in the discretized Brillouin zone, which form a basis of the lattice. 
With this, the Chern number of the nth band is given by
\begin{eqnarray} \label{def_chern_no}
C_n = \frac{1}{2 \pi i} \sum_l  \tilde{F} _{12}({\bf k}_l) ,
\end{eqnarray}
where the sum is over all ${\bf k}_l$ points in a two-dimensional  Brillouin zone.

To compute the chiralities of the Weyl points in the ferri- and ferromagnetic phases, we choose a small two-dimensional sphere
that encloses the given Weyl point and then perform the sum in Eq.~\eqref{def_chern_no} over all ${\bf k}_l$ points on this 
two-dimensional sphere. 
The sign of the resulting integer is  equal to the chirality of the Weyl point. 

The mirror Chern number can be calculated in a similar way, namely, 
as the Chern number in the subspace of occupied states with equal mirror symmetry eigenvalue, see Ref.~\cite{chiu_PRB_17}.

\subsection{Anomalous Hall conductivity}

Weyl points act as sources and sinks of Berry curvature. It is well known that in the presence of    non-zero Berry curvature a non-trivial electronic response can be expected, i.e., an anomalous Hall conductivity. 
That is, the  conductivity tensor $\sigma_{ij}$, defined by the relation ${\bf j}_i = \sigma_{ij} {\bf E}_j$ between the electrical current density $\bf j$ and the electric field $\bf E$, contains a contribution from the anomalous Hall effect. 
This contribution is proportional to the momentum integral of the Berry curvature and can be
written as \cite{RevModPhys.82.1959_NiuBerryPhase} 
\begin{eqnarray}
\sigma_{i j}
&=& 
- 2 \frac{e^2}{\hbar} \int \frac{\text{d}^3k}{(2\pi)^3} \sum_n f(E_n(\bf k)) 
\nonumber \\
& &
 \times \sum_{m \neq n} \frac{
\Im( \bra{n} \frac{\partial H(\bf k)}{\partial k_i} \ket{m} \bra{m} \frac{\partial H(\bf k)}{\partial k_j} \ket{n})
}{(E_n({\bf k}) -  E_m({\bf k}))^2}, 
\label{eq:suppl_exactcurvature}
\end{eqnarray}
where $E_n(\bf k)$ and $\ket{n}$ are the eigenenergy and eigenstate of the $n$-th band, respectively, 
and  $f(E_n(\bf k))$ is the Fermi-Dirac distribution function.
For the numerical evaluation of $\sigma_{ij}$ in the FM phase, the integral in Eq.~\eqref{eq:suppl_exactcurvature}, 
is approximated by its Riemann sum, i.e., 
$\int \frac{\text{d}^3k}{(2\pi)^3}  \to  \sum_{\bf k} \frac{1}{(a N)^3} $, where $a = 5.09 ~\mathrm{\AA}$ 
is the lattice constant of Eu$_3$PbO~\cite{Nuss:dk5032} and $N$ is the number of 
$k$ points per reciprocal lattice direction.

\subsubsection{Symmetries of the anomalous Hall conductivity tensor}

The anomalous Hall conductivity tensor is an antisymmetric matrix of the form
\begin{align} \label{AH_tensor}
\sigma_{\mathrm{general}}
&=
\begin{pmatrix}
0 & \sigma_{xy} &  -\sigma_{zx} \\
 -\sigma_{xy} & 0 & \sigma_{yz}\\
\sigma_{zx} & -\sigma_{yz} &  0 
\end{pmatrix} ,
\end{align}
where the three components $\sigma_{xy}$, $\sigma_{zx}$, and $\sigma_{yz}$
are, in the absence of symmetries, independent of each other.
However, the magnetic space group symmetries put some constrains on this expression. 
To derive these constraints, we first need to consider how the symmetries act on the Berry curvature 
\begin{eqnarray} \label{def_berry_curvature}
\Omega^{n}_{i j}({\bf k})
=
\sum_{m \neq n} \frac{
2 \Im( \bra{n} \frac{\partial H(\bf k)}{\partial k_i} \ket{m} \bra{m} \frac{\partial H(\bf k)}{\partial k_j} \ket{n})
}{(E_n({\bf k}) -  E_m({\bf k}))^2}.  \quad
\end{eqnarray}
First of all, we note that the Berry curvature is even under inversion,  $\Omega^{n}_{i j}({\bf k}) =   \Omega^{n}_{i j}(- {\bf k})$,
but odd under time-reversal symmetry $\Omega^{n}_{i j}({\bf k}) = - \Omega^{n}_{i j}(- {\bf k})$, since complex conjugation 
switches the sign of the imaginary part in the numerator of Eq.~\eqref{def_berry_curvature}. Hence, 
in the paramagnetic phase, where both of these symmetries are present, the Berry curvature is zero. 
In the AFM-I phase the time-reversal symmetry $\mathcal{T}$ is broken.
But there exists a magnetic symmetry  $\widetilde{\mathcal{T}}$ that combines time-reversal 
with a half translation along [111]. Since, the Berry curvature is odd under this symmetry $\widetilde{\mathcal{T}}$, 
the Berry curvature is also vanishing in the AFM-I phase. We conclude that the anomalous Hall conductivity is vanishing
both in the paramagnetic phase and antiferromagnetic phase.

In contrast, the ferri- and ferromagnetic phases exhibit   non-zero anomalous Hall conductivities, as in these phases both 
$\mathcal{T}$ and $\widetilde{\mathcal{T}}$ are broken. We will now focus on the ferromagnetic phase and study how its
symmorphic unitary symmetries constrain the form of the 
anomalous Hall conductivity tensor~\eqref{AH_tensor}. 
A general unitary symmetry $S$ acts on the tight-binding Hamiltonian $H({\bf k})$ of the FM phase as
\begin{eqnarray} \label{unitary_sym_act}
S H({\bf k}) S^{\dag} = H(D_{S} {\bf k}), 
\end{eqnarray} 
where $D_S$ is the momentum space representation of $S$.
From Eq.~\eqref{unitary_sym_act} it follows that for every Bloch eigenstate $\ket{n ( {\bf k} ) }$ with energy $E_n ( {\bf k} )$
there is  a symmetry related eigenstate $\ket{n' ( D_S {\bf k} ) }   = S \ket{n ( {\bf k} ) }$ with the same energy, i.e., 
$E_n ( {\bf k} ) =  E_{n'} ( D_S {\bf k} )$.
Therefore, the band structure is symmetric with respect to the unitary symmetries $S$, and hence the denominator
of the Berry curvature~\eqref{def_berry_curvature} is unchanged under the action of $S$. 
The derivative terms in the numerator of Eq.~\eqref{def_berry_curvature}, on the other hand, are transformed under $S$ as
$S \frac{\partial H(\bf k)}{\partial k_i } S^\dagger = \frac{\partial H(D_{S} \bf k)}{\partial (D_{S} k_i )}$, 
where $\partial / \partial k_i$ and likewise $\partial / \partial (D_{S} k_i )$ are directional derivatives.
With this, we find that 
\begin{eqnarray} \label{eq:supple_symmetryOnCurvature}
&& \bra{n ( {\bf k} ) } \frac{\partial H({\bf k})}{\partial k_\alpha} \ket{m ( {\bf k} )}
=
\bra{n ( \bf{k} )} S^\dagger S \frac{\partial H(\bf k)}{\partial k_\alpha} S^\dagger S \ket{m ( {\bf k} ) }
\nonumber\\
&& \qquad =
\bra{n' ( D_{S} {\bf k} ) }  \frac{\partial H(D_{S} \bf k)}{\partial (D_{S} k_\alpha)} \ket{m' ( D_{S} {\bf k} ) }
\\
&& \qquad =
\bra{n ( D_{S} {\bf k} ) }  \frac{\partial H(D_{S} \bf k)}{\partial (D_{S} k_\alpha)} \ket{m ( D_{S} {\bf k} ) } ,
\nonumber
\end{eqnarray}
where in the last line we have assumed, without loss of generality,  that the Bloch eigenstates $\ket{n ( {\bf k} ) }$ are sorted with increasing eigenenergies.  
Combining Eq.~\eqref{eq:supple_symmetryOnCurvature} with Eq.~\eqref{eq:suppl_exactcurvature}, 
we see that the summands in Eq.~\eqref{eq:suppl_exactcurvature} can be grouped into symmetry related pairs, namely  
 $\Omega^n_{ij} ( {\bf k} )$  and $\Omega^n_{ij} ( D_S {\bf k})$, whose contributions 
 differ only by directional derivatives $\partial / \partial k_{i/j}$ and $\partial / \partial ( D_S k_{i/j} )$.
 If $S$ is a mirror symmetry with mirror plane perpendicular to the main axes (or a two-fold rotation symmetry about a main axis), we find that
 $D_S k_{i/j} = \pm k_{i/j}$, where the sign depends on whether $k_{i/j}$ is perpendicular or parallel to the mirror plane. 
 Since there are two derivative terms in Eq.~\eqref{def_berry_curvature}, we have
 $\Omega^{n}_{ij}({\bf k}) = - \Omega^{n}_{ij}(D_{S} {\bf k})$, when, for example, 
 $D_S k_i = + k_i$ but $D_S k_j = - k_j$, and hence the corresponding
 component of the anomalous Hall conductivity tensor is vanishing. 
 Similar arguments can also be constructed for the three-fold and four-fold rotation symmetries of the FM phase.

Applying the above symmetry analysis to the FM phase with magnetization orientations [100], [110], and [111], we
find that conductivity tensors are of the form 
\begin{align}
\sigma_{\mathrm{FM100}}
&=
\begin{pmatrix}
0 & 0 & 0 \\
0 & 0 & \Sigma\\
0 & -\Sigma &  0 
\end{pmatrix} ,
\\
\sigma_{\mathrm{FM110}}
&=
\begin{pmatrix}
0 & 0&  -\Sigma' \\
 0 & 0 & \Sigma'\\
\Sigma' & -\Sigma' &  0 
\end{pmatrix} ,
\\
\sigma_{\mathrm{FM111}}
&=
\begin{pmatrix}
0 & \Sigma'' &  -\Sigma'' \\
 -\Sigma'' & 0 &\Sigma''\\
\Sigma'' & -\Sigma'' &  0 
\end{pmatrix}  ,
\end{align} 
for some nonzero $\Sigma, \Sigma', \Sigma'' \in \mathbb{R}$.
We observe that for these high-symmetry magnetization directions,
the three components of the conductivity tensor~\eqref{AH_tensor}
are dependent on each other.

\subsubsection{Contributions of the different Weyl points to the anomalous Hall conductivity}

\begin{table}[tbh]
\centering
\begin{ruledtabular}
\begin{tabular}{ccclcc}
phase & position    &  E (eV)  & name & chirality & $\sigma_{yz}$ \\ 
\hline
FM [100]& {\footnotesize $(0.18, 0, 0)$ }& -0.16 & W2 & 1 & -274 \\
FM [100] &{\footnotesize $(0.12, 0, 0)$ } & -0.33 & W3 & -1 & 183 \\
\hline
FM [110] & {\footnotesize $(0.23, 0.015, 0)$} & 0.06 &  W1 & -1 & 352 \\
FM [110]& {\footnotesize $(0.19, -0.001, 0)$ }& -0.12 &  W2 & 1 & -289 \\
FM [110] &{\footnotesize $(0.12, 0, 0) $ } & -0.32 &  W3 & -1 & 183 \\
\hline
FM [111] & {\footnotesize $(0.23, 0.008, 0.008)$} & 0.05 & W1 & -1 & 351 \\
FM [111] & {\footnotesize $(0.2,-0.001,-0.001)$ }& -0.09 & W2 & 1 & -303 \\
FM [111] &{\footnotesize $(0.13, 0, 0)$ } & -0.31 &  W3 & -1 & 198 
\end{tabular}
\end{ruledtabular}
\caption{
\textbf{Anomalous Hall conductivity of Weyl points. }
This table lists the positions and energies of a selection of Weyl points in the first Brillouin zone (BZ)
for the ferromagnetic phase with magnetization in [100], [110], and [111]   direction (FM [100], FM [110], and FM [111], respectively).
 The positions of the band crossings ${\bf k} = ( k_x, k_y, k_z)$ are given in 
units of $2 \pi/ a_{i}$, where $a_i$ denotes the lattice constant of the respective real space direction. All energies are given in eV relative to the Fermi energy. 
The names of band crossings are indicated in the fourth column, while the fifth column states the chirality for the Weyl point at the given position. The last column gives the respective contribution to the anomalous Hall conductivity in units of $(\Omega \mathrm{ cm})^{-1}$.
\label{table_WeylpointAHE} 
}
\end{table} 

It is known from field theoretical considerations as well as calculations in lattice  systems that the 
conductivity of a Weyl semimetal at half-filling is given by~\cite{2013_prb_Goswami_WeylConductivity} 
\begin{align}
\sigma_{\alpha \beta} = - \frac{e^2}{\hbar} \epsilon_{\alpha \beta \gamma} \frac{b_\gamma}{2 \pi^2},
\label{Eq:CondFromSeparation}
\end{align}
where $2 b_\gamma$ is a component of the vector that connects the negative chirality Weyl point  to the one with positive chirality. While Eq. (\ref{Eq:CondFromSeparation}) allows us to estimate the largest possible contribution of a pair of Weyl points to the conductivity, it gives no further information about how the conductivity depends on the chemical potential,
as shown by the curves in Fig.~\ref{mFig4} of the main text.
The shape and width of the extrema of these curves depend on the details of the band structure, in particular, on how strongly the bands disperse. 

Nevertheless, the location of the extrema in Fig.~\ref{mFig4}  can be explained to a large extent by the presence of Weyl points. In Table~\ref{table_WeylpointAHE} we list the 
contributions to the conductivity  from the Weyl points W1, W2, and W3, as given by Eq.~\eqref{Eq:CondFromSeparation}. By comparison to Fig.~\ref{mFig4} we can conclude that the different Weyl point contributions overlap and partially cancel. We observe that the conductivity peak of W2 spreads over a larger energy range than W1 and W3. 
We also note that the large positive contribution of W1 distinguishes the conductivity of FM [110] and FM [111] from FM [100].

\subsection{Surface States and Spin Texture} 

To compute the surface state spectra, we perform a Fourier transform of the Hamiltonian $H ({\bf k} )$
in the direction perpendicular to the surface, say, the $z$ direction. 
This yields a Hamiltonian $H (k_x, k_y, z )$ that depends on two momenta,
$k_x$  and $k_y$, and on one real space coordinate $z$.
We then numerically diagonalize the Hamiltonian $H (k_x, k_y, z )$ with open boundary conditions along the $z$ direction,
to obtain the eigenstates $\psi_n ( k_x, k_y)$ and energies $E_n (k_x, k_y)$. 
From these we can compute the surface density of states, which is given by 
\begin{align} \label{def_eq_SDOS}
\rho(\omega, k_x, k_y, z) &= -\frac{1}{N} \mathrm{Im} \sum_n \frac{\psi_n^\dagger(k_x, k_y, z) \psi_n(k_x, k_y, z)}{\omega + i \Gamma/N - E_n(k_x, k_y)},
\nonumber\\
\rho(\omega, k_x, k_y) &= \sum_{z \in \mathrm{surface} } \rho(\omega, k_x, k_y, z) ,
\end{align}
where $N$ denotes the number of layers perpendicular to the $z$ direction.
Here, $\Gamma$ is a phenomenological broadening factor that takes into account the
effects of finite temperature and disorder. For our numerical calculations
we have chosen  $\Gamma = 1$ and $N=70$.  
The sum in the second line of Eq.~\eqref{def_eq_SDOS} is
taken over the first five surface layers.

The spin polarization of the surface states is calculated very similarly by considering the weighted expectation value
\begin{align}
\rho(\omega, k_x, k_y)_{\alpha} &= -\frac{1}{N} \mathrm{Im} \sum_{n, z}  \frac{\psi_n^\dagger(k_x, k_y, z) \tilde{\sigma}_{\alpha} \psi_n(k_x, k_y, z)}{\omega + i \Gamma/N - E_n(k_x, k_y)},
\end{align} 
where $\tilde{\sigma}_{\alpha}$ is the spin matrix of the 18 band basis for the $\alpha$ direction, e.g., for $\alpha = x$ a tensor product between unity operators in orbital space and the Pauli matrix $\sigma_x$.

 \clearpage
 
%
%
%

\end{document}